\documentclass[11pt]{article}%
\usepackage{amsmath}
\usepackage{graphicx}
\usepackage{array}
\usepackage{booktabs}
\usepackage{subfigure}
\usepackage{float}
\usepackage{url}
\usepackage{color}
\usepackage[ruled,linesnumbered]{algorithm2e}%
\usepackage{amsfonts}%
\usepackage{amssymb}

\newcommand{\eqnb}{\begin{equation}}
\newcommand{\eqne}{\end{equation}}

\newtheorem{The}{Theorem}

\newtheorem{Rem}{Remark}

\begin{document}

\title{Performance and Reliability Analysis for Practical Byzantine Fault Tolerance with Repairable Voting Nodes}
\author{Yan-Xia Chang$^{1}$, Qing Wang$^{2}$,  Quan-Lin Li$^{1}$\thanks{Corresponding author: Q. L. Li
(liquanlin@bjut.edu.cn)}, Yaqian Ma$^{1}$\\$^{1}$School of Economics and Management\\Beijing University of Technology, Beijing 100124, China\\$^{2}$Monash Business School\\Monash University, Melbourne, Australia}
\maketitle
\footnotetext{ This work has been submitted to the IEEE for possible publication. Copyright may be transferred without notice, after which this version may no longer be accessible.  }

\begin{abstract}
\vskip         0.5cm

The practical Byzantine fault tolerant (PBFT) consensus protocol is one of the basic consensus protocols in the development of blockchain technology. At the same time, the PBFT consensus protocol forms a basis for some other important BFT consensus protocols, such as Tendermint, Streamlet, HotStuff, and LibraBFT. In general, the voting nodes may always fail so that they can leave the PBFT-based blockchain system in a random time interval, making the number of timely available voting nodes uncertain. Thus, this uncertainty leads to the analysis of the PBFT-based blockchain systems with repairable voting nodes being more challenging. In this paper, we develop a novel PBFT consensus protocol with repairable voting nodes and study such a new blockchain system using a multi-dimensional Markov process and the first passage time method. Based on this, we provide performance and reliability analysis, including throughput, availability, and reliability, for the new PBFT-based blockchain system with repairable voting nodes. Furthermore, we provide an approximate algorithm for computing the throughput of the new PBFT-based blockchain system. We employ numerical examples to demonstrate the validity of our theoretical results and illustrate how the key system parameters influence performance measures of the PBFT-based blockchain system with repairable voting nodes. We hope the methodology and results developed in this paper will stimulate future research endeavors and open up new research trajectories in this field.


\end{abstract}

\section{Introduction}

As one of the fundamental problems in computer science, the consensus problem in distributed computing field was first raised by Pease, Shostak, and Lamport \cite{Pease:1980} in 1980, and the later was named ``The Byzantine General problem" by \cite{Lam:1982, Lam:1983}. Along with computer development, the consensus problem has received more attention over the last decade due to its extensive applications to a new distributed system called ``blockchain" such as Bitcoin \cite{Nak:2008} and Ethereum \cite{But:2013}. Driven by the opportunities in the concept of cryptocurrencies and blockchain, a variety of consensus has emerged \cite{Ban:2017, Cac:2017}, where BFTs or PBFTs play a crucial role in the evolution of the blockchain ledger consensus.

In contrast to public blockchains such as Bitcoin \cite{Nak:2008} and Ethereum 1.0 \cite{But:2013}, which are based on Proof-of-Work (PoW) consensus protocol, PBFT-based blockchain system tends to have only a subset of participating nodes running the PBFT protocol. Those nodes can be tightly controlled by the membership, and once more than 2/3 of the voting nodes have approved a proposal proposed by the primary node, a round of voting consensus is reached. For such a PBFT-based blockchain system that allows less than 1/3 of voting nodes to fail due to some reasons, it can work correctly even under malicious attacks, software errors, operator mistakes, and so on. Based on this fault-tolerant capability, the PBFT-based blockchain has some advantages, such as access control, low energy consumption, high throughput, fast consensus speed, and high scalability. For more details, readers may refer to several survey papers by, for example, Correia et al. \cite{Correia:2011}, Vukoli{\'c} \cite{Vuko:2016}, Gramoli \cite{Gramo:2020}, Berger and Reiser \cite{Berger:2018}, Gupta et al. \cite{Gupta:2019}, Stifter et al. \cite{Stif:2019}, Alqahtani and Demirbas \cite{Alqah:2021}, Zheng and Feng \cite{Zheng:2021}, Gan et al. \cite{Gan:2021} and others.

Although PBFT-based blockchain systems offer the above advantages, it also has some limitations. For example, an ordinary PBFT-based blockchain has poor system flexibility. In other words, the voting nodes cannot freely join or exit the PBFT network. Therefore, it is impossible for a more general PBFT-based blockchain system to take into account the behavior of failed nodes who leave the PBFT-based blockchain system or of repaired nodes who reconnect to the PBFT network. This limitation goes against the principle that all the nodes of the PBFT-based voting network have the right to vote on a proposal or suggestion over time. Also, if more than 1/3 of nodes fail, and the failed nodes have not been repaired in a timely manner, then the PBFT-based blockchain system will never reach a consensus. Obviously, this limitation is also harmful to the liveness, availability, and security of the PBFT-based blockchain system. Thus, it is necessary to consider a new PBFT consensus protocol, in which the failed nodes may leave the PBFT network, while the repaired nodes can re-enter the PBFT network. By introducing a failure and repair process, the nodes that fail and repair can change the number of voting nodes in the PBFT network, allowing us to observe the availability and security of the PBFT-based blockchain system in a timely manner. In addition, the inclusion of repaired nodes allows the PBFT network to perform the voting consensus, even if the PBFT-based blockchain system is on the verge of becoming unavailable due to the larger number of failed nodes in the system.

In this paper, we focus on the PBFT-based blockchain system with repairable voting nodes and provide performance and reliability analysis for such a new blockchain system using a multi-dimensional Markov process and the first passage time method. Meanwhile, we develop some key performance and reliability measures for the PBFT-based blockchain system with repairable voting nodes, including throughput, availability, and reliability. These are all important measures to describe and evaluate blockchain systems. Note that in our paper, the Markov processes and queueing theory play a key role in the study of PBFT-based blockchain systems. A growing body of literature has applied Markov processes and queueing theory to study blockchain systems. For example, a simple Markov chain by Eyal and Sirer \cite{Eyal:2014} , a Markov queueing model by Li et al. \cite{Li:2018, Li:2019}, a two-dimensional Markov process by G\"{o}bel et al. \cite{Gobel:2016} and Javier and Fralix \cite{Jav:2020}, a pyramid Markov process by Li et al. \cite{Li:2020}, a Markov process of DAG-based blockchain by Song et al. \cite{Song:2022}, a Markov process of PBFT-based blockchain by Ma et al. \cite{Ma:2022}, and a Markov queueing model of dynamic PBFT-based blockchain system by Chang et al. \cite{Chang:2022} and so on. From those works, it becomes evident that Markov processes and queueing theory can well describe some different consensus processes of blockchain systems, and they can have universality and superiority in evaluating the performance measures of blockchain systems.

This paper is closely related to Ma et al. \cite{Ma:2022}, Chang et al. \cite{Chang:2022}, Nischwitz et al. \cite{Nis:2021}, and Hao et al. \cite{Hao:2018}, which provided performance analysis for the PBFT-based blockchain systems by using Markov processes or probabilistic models. Hao et al. \cite{Hao:2018} presented a dynamic PBFT-based blockchain system with an uncertain number of voting nodes, in which some nodes may enter or leave the PBFT network by means of the JOIN and EXIT protocols. Similarly, Chang et al. \cite{Chang:2022} described and analyzed the dynamic (entering and leaving) behavior of some nodes in the PBFT network by using the Markov process theory and an approximate queueing system. Compared with Hao et al. \cite{Hao:2018} and Chang et al. \cite{Chang:2022}, this paper uses a three-dimensional Markov process to describe and analyze the behavior of failed nodes who leave the PBFT network and of repaired nodes who re-enter the PBFT network while the total number of voting nodes (including the failed nodes) is fixed. Based on this, we obtain the probability distributions of block-generated and orphan-block-generated are of phase-type, and then we analyze the throughput of the PBFT-based blockchain system. Differently with this paper, Chang et al. \cite{Chang:2022} provided a steady-state rate approximate method such that the block generation time and the orphan block generation time were exponential, and then they set up an approximate queueing model to compute the throughput of the PBFT-based blockchain system. Ma et al. \cite{Ma:2022} adopted a two-dimensional Markov process to describe the voting process, which can be considered a simple and special case of our works. Nischwitz et al. \cite{Nis:2021} used a probabilistic model to evaluate BFT protocols in the presence of dynamic link and crash failures, which is different from the methodology of this paper.

Based on the above analysis, the main contributions of this paper are summarized as follows:

\begin{itemize}
\item[1.] We propose a new PBFT consensus protocol to generalize the ordinary PBFT consensus protocol, in which the failed nodes may leave the PBFT network and some repaired nodes can enter the PBFT network again, so the number of available voting nodes is variable over time. Based on this, we propose a large-scale Markov modeling technique to analyze the performance and reliability of the new PBFT-based blockchain system with repairable voting nodes.
\item[2.] We use PH distributions of finite size to express the probability distributions of block-generated and orphan-block-generated times. Then we use the two PH distributions to set up a more realistic queueing model $\mathrm{{M}} \oplus\mathrm{{P}}{\mathrm{{H}}^{b}}/\mathrm{{P}}{\mathrm{{H}}^{b}}/1$ to compute the throughput of the PBFT-based blockchain system with repairable voting nodes, and provide an approximate algorithm for computing the throughput of the PBFT-based blockchain system.
\item[3.] We provide reliability analysis for the PBFT-based blockchain systems with repairable voting nodes and expressions for availability, reliability, and the average time before the first failure.
\item[4.] We use numerical examples to validate our theoretical results and demonstrate the impact of key parameters on performance measures of the PBFT-based blockchain system with repairable voting nodes.
\end{itemize}

The structure of this paper is organized as follows. Section \ref{sec:review} surveys related literature on the performance evaluation of PBFT-based blockchain systems. Section \ref{sec:model descrition} describes a new PBFT-based blockchain system with repairable voting nodes. Section \ref{sec:distribution} analyzes the probability distributions of
block-generated and orphan-block-generated times by means of two phase-type distributions of finite sizes. Section \ref{sec:blockchain system} introduces queueing model $\mathrm{{M}} \oplus\mathrm{{P}}{\mathrm{{H}}^{b}}/\mathrm{{P}}{\mathrm{{H}}^{b}}/1$ to provide performance analysis for the PBFT-based blockchain system with repairable voting nodes. Section \ref{sec:security and reliability} sets up two new Markov processes to analyze the reliability of the PBFT-based blockchain systems with repairable voting nodes. Section \ref{sec:numberical analysis} uses some numerical examples to indicate how the key parameters influence the performance measures of the PBFT-based blockchain system with repairable voting nodes. Section \ref{sec:concluding} gives some concluding remarks. Finally, two appendixes provide the non-zero matrix elements or blocks in two important infinitesimal generators.

\section{Literature Review}\label{sec:review}

In this section, we divide current research into three literature streams: The first is the research on the development and optimization of the PBFT consensus protocol. The second is on analytical models to evaluate the performance of PBFT-based blockchain systems. The third is on simulation models to study the performance of PBFT-based blockchain systems.

\textbf{(a) Research on the development and optimization of the PBFT consensus protocol}

Since Lamport, Shostak, and Pease \cite{Pease:1980} proposed the Byzantine General problem in 1982, there has been a growing demand to deploy BFT and/or its variations into practical applications. This demand has driven the continuous design and improvement of BFT protocol as well as extensive research utilizing the BFT algorithm. Important examples include Schlichting and Schneider \cite{Schlichting:1983}, Reischuk \cite{Rei:1985}, Martin and Alvisi \cite{Mar:2006}, Veronese et al. \cite{Veron:2011}, Malkhi et al. \cite{Malk:2019} and so on. However, many assumptions of some early studies were too idealistic to be realized in reality. To make the BFT protocol more suitable for practical applications, Castro and Liskov \cite{Castro:1999, Cas:2002} improved the BFT and proposed the PBFT protocol, which is the first practical solution to the Byzantine problem and can work in asynchronous environments. Henceforward, some researchers have further developed the BFT or PBFT, effectively improving the performance of the BFT-based protocol. These solutions include but are not limited to dividing replicas into groups \cite{Yu:2020, Thai:2019}, introducing hierarchical structure \cite{Thai:2019}, simplifying the PBFT process \cite{Crain:2018, Chen:2020, LiP:2020, LiY:2019}, introducing credit mechanism \cite{Yu:2020, Chen:2020, Wang:2019, Wang:2020, Tong:2019, Gao:2019}, using multiple consensus in different cases \cite{Yu:2020, Sakho:2020, Chris:2019}, and introducing geography factors \cite{Gueta:2019, Lao:2020}.

Inspired by these solutions, this paper proposes a new PBFT consensus protocol with repairable voting nodes, which is different from the static PBFT protocol described in the literature above. In such a blockchain system based on this new protocol, failed nodes may leave the PBFT network, and some repaired nodes can enter the PBFT network again. By introducing the fail and repair processes, failed and repaired nodes can change the number of voting nodes in the PBFT network, thus transforming the PBFT protocol from a completely closed environment to an open one. While we hope that the new proposed PBFT can inherit the advantages of the previous PBFT, we also hope that this PBFT-based blockchain system with repairable voting nodes can guarantee liveness, availability, and security.

\textbf{(b) Research on the performance evaluation based on analytical models  }

As one of the core components of a blockchain system, the PBFT consensus protocol directly affects the overall throughput, delay, reliability, and fault tolerance of the blockchain system, and determines the efficiency and scalability of a blockchain system to a large extent. Up to now, BFT or PBFT protocol has become the most basic one in all the blockchain consensus mechanisms and has been used in many areas, such as the Internet of Things (IoT) \cite{Lao:2020, Meshch:2021, Yuan:2021}, Internet of vehicles (IoV) \cite{Hu:2019}, cloud computing \cite{Lim:2014}, energy trading \cite{Shei:2019}, and many other fields. As more PBFT-based protocols are applied to reality, evaluating the performance of these blockchain systems becomes increasingly important, which is the first important thing we did after proposing the new PBFT consensus protocol with repairable voting nodes. Readers may refer to the surveys of Fan et al. \cite{Fan:2020} and Dabbagh et al. \cite{Dabba:2021} for the methods used in recent years to evaluate the performance of PBFT-based blockchain systems.

As we all know, the analytical model mainly starts from the blockchain system itself, which helps to deeply understand the influencing factors, evolution, and development of the blockchain system. Based on this characteristic, these analytical methods can evaluate, interpret, and predict the behavior of PBFT-based blockchain systems. Among studies using analytical models, the closest works to ours are Chang et al. \cite{Chang:2022}, Ma et al. \cite{Ma:2022}, Nischwitz et al. \cite{Nis:2021}, and Hao et al. \cite{Hao:2018}, in which Markov processes or probabilistic models were introduced to evaluate the performance of the PBFT-based blockchain systems. Chang et al. \cite{Chang:2022} set up a large-scale Markov process to describe a novel dynamic PBFT where the voting nodes may always leave the network while new nodes may also enter the network and provided an effective computational method for the throughput. Ma et al. \cite{Ma:2022} described a simple stochastic performance model to analyze the voting process of the PBFT-based blockchain system. Nischwitz et al. \cite{Nis:2021} used a probabilistic model to evaluate BFT protocols in the presence of dynamic links and crash failures. Hao et al. \cite{Hao:2018} proposed a dynamic PBFT system from the perspective of protocols and used a probabilistic model to calculate the probability that the client gets the correct message. Different from these articles, we not only use the Markov process to analyze the throughput but, in particular, we also analyze the reliability of the PBFT-based blockchain system with repairable voting nodes.

In addition, many researchers have made unremitting efforts to evaluate the performance of PBFT-based blockchain systems from different angles and using various analysis methods. For example, Hao et al. \cite{HaoY:2018} proposed a method to evaluate the performance of PBFT consensus in Hyperledger and showed that PBFT consistently outperforms PoW in terms of latency and throughput under varying workloads. Sukhwani et al. \cite{Sukh:2017} modeled the PBFT consensus process using Stochastic Reward Nets (SRN) to compute the mean time to complete consensus for networks of up to 100 peers. Lor{\"u}nser \cite{Loru:2022} presented a performance model for PBFT that specifically considered the impact of unreliable channels and the use of different transport protocols over them. Pongnumkul et al. \cite{Pong:2017} developed a method to evaluate the Hyperledger Fabric and Ethereum and showed that Hyperledger Fabric consistently outperforms Ethereum across all evaluation metrics (execution time, latency, and throughput).

\textbf{(c) Research on the performance evaluation based on simulation models }

As one of the commonly used methods to evaluate the performance of blockchain, simulation models are primarily used to simulate or approximate the behavior and development of practical blockchain systems with computers in situations where the description of analytical models is not possible or where it is challenging to set up an analytical model. Based on the simulation models, Meshcheryakov et al. \cite{Meshch:2021} argued the effectiveness of the PBFT consensus algorithm for its implementation on constrained IoT devices, simulated the main distributed ledger scenarios using PBFT, and evaluated the blockchain system performance. Monrat et al. \cite{Monrat:2020} provided a performance and scalability analysis of popular private blockchain platforms by varying the workloads and determining the performance evaluation metrics such as throughput and network latency. Ahmad et al. \cite{Ahmad:2021} developed a blockchain test platform to execute and test the latency and throughput of blockchain systems, including PBFT, PoW, Proof of Equity (PoS), Proof of Elapsed Time (PoET), and Clique. Zheng et al. \cite{Zheng:2018} used continuous-time Markov chain (CTMC) models to simulate the time response of a PBFT-based healthcare blockchain network.

From the above literature streams, existing relevant literature has studied the performance of PBFT-based blockchain systems from three aspects: protocol, analytical models, and simulation models. The former focuses on improving the PBFT protocol to evaluate and improve the performance of the PBFT-based blockchain systems from the protocol itself, while the latter two mainly evaluate the performance of the PBFT-based blockchain system from the operating mechanism of the blockchain itself. However, few studies have used the Markov process theory to analyze the performance and reliability of PBFT-based blockchain systems with repairable voting nodes. Motivated by the scarcity of research on the performance and reliability of PBFT-based blockchain systems with repairable voting nodes, we are compelled to develop this topic in this paper.

\section{Model Description}

\label{sec:model descrition} In this section, we provide a detailed model
description for a PBFT-based blockchain system with repairable voting nodes. Also, we give
mathematical notation, random factors, and necessary parameters used in our
later study.

\textbf{(1) The failure process of voting nodes:} We assume that each voting
node can fail in the PBFT-based blockchain system, and the lifetime of the voting node is
exponential with mean $1/\theta>0$. Once a voting node fails, it cannot
handle any other work or task until it is repaired well to enter the working state again.

\textbf{(2) The repair process of voting nodes:} Once any voting node fails, it
immediately enters a repair state. We assume that the repair time of the
failed voting node is exponential with mean $1/\mu>0$.

\textbf{(3) The total number of voting nodes:} For the convenience of
analysis, we assume that the total number of voting nodes in the PBFT network
is a fixed value $N=3n+1$, where $n$ is a given positive integer.

\textbf{(4) The voting process:} We assume that the voting time of each node
is exponential with mean $1/\gamma>0$. Additionally, we assume that each node can have only one voting chance with a voting period.

\textbf{(5) The probability that a transaction package is either approved or
disapproved:} Using the law of large numbers from some voting statistics, we assume that the
probability that each voting node approves a transaction package is $p$;
while the probability that each voting node disapproves a transaction package is $q=1-p$.

\textbf{(6) The judgment of voting result:} We denote by $N(t)$, $M(t)$, and
$K(t)$ the number of voting nodes that approve the transaction package, the
number of voting nodes that disapprove the transaction package, and the number
of failed nodes at time $t>0$, respectively. Based on this assumption, we judge the voting results in a round of voting process as follows:

\textbf{(a)} If $N(t)=2n+1$, then the number of approval votes is more than
2/3 of the total number of voting nodes. This means that a transaction package
can be determined as a block so that the block can be pegged on the blockchain.

\textbf{(b)} If $M(t)+K(t)=n+1$, then the number of approval votes is
less than 2/3 of the total number of voting nodes. This means that a
transaction package can be determined as an orphan block so that the orphan block must be
rolled back to the transaction pool to be dealt with again.

\textbf{(7) The times of pegging a block and rolling back an orphan block:} Note that
the processes of pegging a block and rolling back an orphan block are all performed in
the PBFT-based blockchain network. Thus, we regard the block-pegging time and the orphan block rolling-back time as identical, and we assume that the
block-pegging time and the orphan block rolling-back time are exponential with the mean
$1/\beta$.

\textbf{(8) Independence:} We assume that all random variables defined
above are independent of each other.

\begin{Rem} From Assumption (6), there exist three different cases in which a transaction package is determined as an orphan block: \textbf{(1)} $M(t)=n+1$,
$K(t)=0$; \textbf{(2)} $M(t)=0$, $K(t)=n+1$; \textbf{(3)} $M(t)+K(t)=n+1$ for
$M(t)\ge1$ and $K(t)\ge1$.
\end{Rem}

\begin{Rem}
In the process of PBFT voting, once a transaction package is determined as a block (i.e.,
$N(t)=2n+1$), it is not necessary
to further perform any subsequent process such that $N(t)>2n+1$.
Similarly, once a transaction package is determined as an orphan block (i.e.,
$M(t)+K(t)=n+1$), it is not necessary to further perform any subsequent process such that $M(t)+K(t)>n+1$.
\end{Rem}

\section{Block- and Orphan-Block-Generated Time Distributions}\label{sec:distribution}
In this section, we express the probability distributions of block- and orphan-block-generated times by means of the phase-type distributions of
finite sizes.

Note that $N(t)$, $M(t)$, and $K(t)$ denote the number of voting nodes that
approve the transaction package, the number of voting nodes that disapprove
the transaction package, and the number of failed nodes at time
$t\ge0$, respectively. It is easy to see that $\left\{  {\left(
{N(t),M(t),K(t)} \right)  :t \ge0} \right\}  $ is a three-dimensional
continuous-time Markov process whose state space is given by
\[
\Omega= \mathop    \cup\limits_{k = 0}^{2n + 1} \mathrm{{ Level }}\; k,
\]
where for $0 \le k \le2n$,
\begin{align*}
\mathrm{{Level }} \; k =  &  \left\{  {(k,0,0),(k,0,1), \ldots,(k,0,n -
1),(k,0,n),(k,0,n + 1)} \right.  ;\\
&  (k,1,0),(k,1,1), \ldots,(k,1,n - 1),(k,1,n);\\
&  (k,2,0),(k,2,1), \ldots,(k,2,n - 1); \ldots;\\
&  \left.  {(k,n + 1,0)} \right\}  ,
\end{align*}
and for $k=2n+1$,
\begin{align*}
\mathrm{{Level }}\;k =  &  \left\{  {(k,0,0),(k,0,1), \ldots,(k,0,n -
2),(k,0,n - 1),(k,0,n)} \right.  ;\\
&  (k,1,0),(k,1,1), \ldots,(k,1,n - 2),(k,1,n - 1);\\
&  (k,2,0),(k,2,1), \ldots,(k,2,n - 2); \ldots;\\
&  \left.  {(k,n,0)} \right\}  .
\end{align*}

\begin{figure}[ptbh]
\centering        \includegraphics[width=12cm]{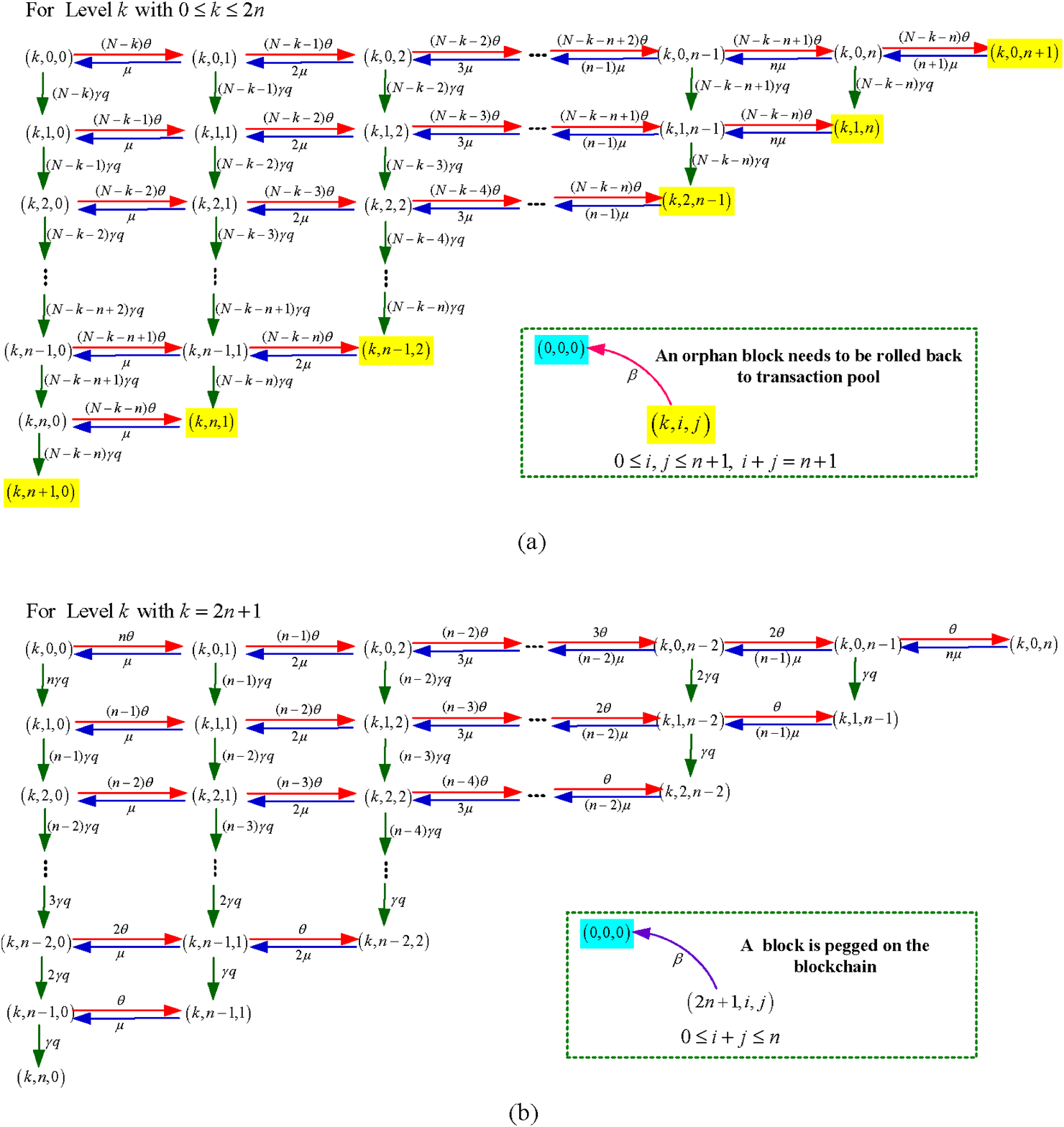}  \caption{The state transition relations of the Markov process within one level.}%
\label{figure:Fig-1}%
\end{figure}

In Fig. 1(a), a yellow state $(k,i,j)$ with $i+j=n+1$ denotes that the
transaction package is determined as an orphan block, and it needs to be rolled back to
the transaction pool. In Fig. 1(b), each state in Level $(2n+1)$ is
determined as a block, and it can be pegged on the blockchain. Once a transaction
package is determined as either a block or an orphan block, this round of voting process
is over. Then, the blockchain system enters the state $(0,0,0)$ such that a new round of
voting process begins again.

\begin{figure}[ptbh]
\centering        \includegraphics[width=8cm]{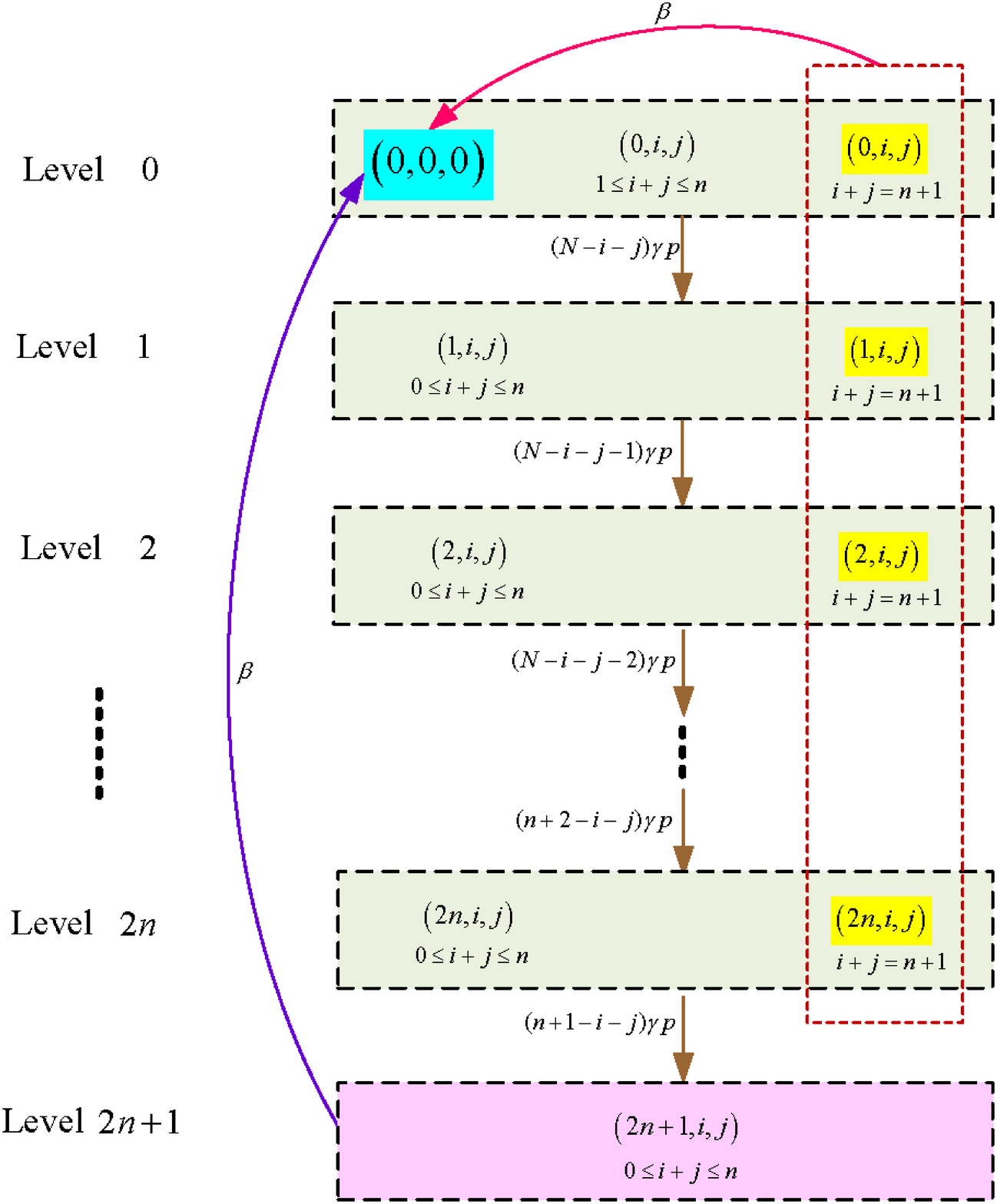}  \caption{The state
transition relations of the Markov process among multiple levels.}%
\label{figure:Fig-2}%
\end{figure}

Using the Markov process $\left\{  {\left(  {N(t),M(t),K(t)} \right)  :t
\ge0} \right\}  $, it is easy to see that the transaction package becomes a block at each of
states $(2n + 1,i,j)$ for $0 \le i,j \le n$; while the transaction package becomes an
orphan block at states $(k,i,j)$ with $0 \le k \le2n$, $0 \le i,j \le n + 1$, $i +
j = n + 1$. Also, the events that determine a transaction package as either a block or an orphan block are mutually exclusive. Therefore,
we can analyze the probability distribution that a transaction package is generated as either a block
or an orphan block.

\subsection{The block-generated time distribution} \label{subsec:block-generated time}

\begin{figure}[ptbh]
\centering        \includegraphics[width=13cm]{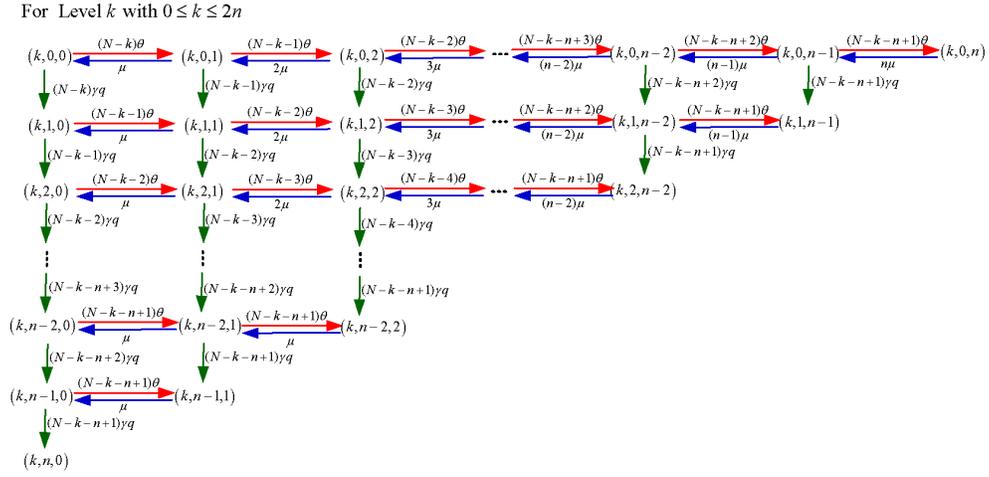}  \caption{The state
transition relations of the Markov process $\Psi$: Part one.}%
\label{figure:Fig-3}%
\end{figure}

\begin{figure}[ptbh]
\centering        \includegraphics[width=5cm]{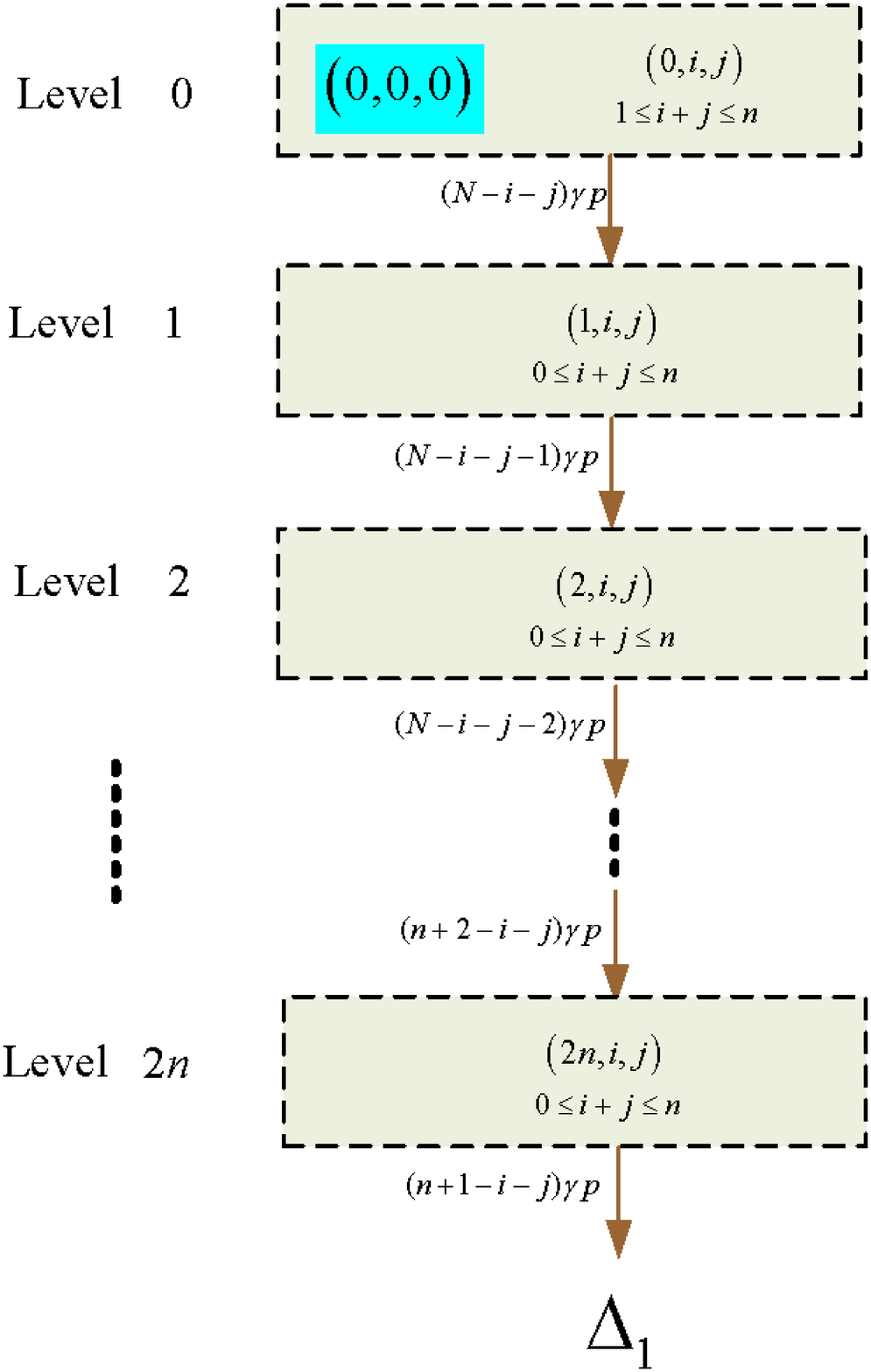}  \caption{The state
transition relations of the Markov process $\Psi$: Part two.}%
\label{figure:Fig-4}%
\end{figure}

To analyze the block-generated time
distribution, we set up a new Markov process with an absorption state
${\Delta_{1}}$. To do this, all the states in the set
\[
\left\{  {(2n+1,i,j):0 \le i,j \le n} \right\}
\]
are regarded as an absorption state ${\Delta_{1}}$. In this case, the Markov process
$\left\{  {\left(  {N(t),M(t),K(t)} \right)  :t \ge0} \right\}  $ operates on
a new state space with the absorption state ${\Delta_{1}}$ as follows:
\[
\left\{  {{\Delta_{1}}} \right\}  \cup\left\{  {(k,i,j):k =  {0,1,2,
\ldots,2n}  ,0 \le i + j \le n} \right\}  .
\]
At the same time, the state transition relations of the Markov process $\left\{  {\left(
{N(t),M(t),K(t)} \right)  :t \ge0} \right\}  $ with the absorption state
${\Delta_{1}}$ are depicted in Figures \ref{figure:Fig-3} and
\ref{figure:Fig-4}. Also, its infinitesimal generator is given by
\[
\Psi=\left(  {%
\begin{array}
[c]{cc}%
0 & 0\\
{{T^{0}}} & T
\end{array}
}\right)  ,
\]
where, ${T^{0}}+T\mathbf{{e}}=0$,
\[
{T^{0}}=\left(  {%
\begin{array}
[c]{c}%
0\\
\vdots\\
0\\
{T_{2n}^{0}}%
\end{array}
}\right)  ,T_{2n}^{0}=\left(  {%
\begin{array}
[c]{c}%
{{L_{0}}}\\
{{L_{1}}}\\
\vdots\\
{{L_{n}}}%
\end{array}
}\right)  ,
\]%
\[
{L_{i}}={\left(  {%
\begin{array}
[c]{c}%
{\left(  {n+1-i}\right)  \gamma p}\\
{\left(  {n-i}\right)  \gamma p}\\
\vdots\\
{\gamma p}%
\end{array}
}\right)  _{(n+1-i)\times1}},0\leq i\leq n;
\]%
\[
T=\left(  {%
\begin{array}
[c]{ccccc}%
{{T_{0,0}}} & {{T_{0,1}}} &  &  & \\
& {{T_{1,1}}} & {{T_{1,2}}} &  & \\
&  & \ddots & \ddots & \\
&  &  & {{T_{2n-1,2n-1}}} & {{T_{2n-1,2n}}}\\
&  &  &  & {{T_{2n,2n}}}%
\end{array}
}\right)  ,
\]
for $0\leq k\leq2n-1$,
\[
{T_{k,k+1}}=\left(  {%
\begin{array}
[c]{cccc}%
{{F_{0,0}^{(k)}}} &  &  & \\
& {{F_{1,1}^{(k)}}} &  & \\
&  & \ddots & \\
&  &  & {{F_{n,n}^{(k)}}}%
\end{array}
}\right)  ,
\]
for $0\leq i\leq n$,
\[
{F_{i,i}^{(k)}}={\left(  {%
\begin{array}
[c]{cccc}%
{(N-k-i)\gamma p} &  &  & \\
& {(N-k-i-1)\gamma p} &  & \\
&  & \ddots & \\
&  &  & {(N-k-n)\gamma p}%
\end{array}
}\right)  _{(n+1-i)\times(n+1-i)}};
\]
for $0\leq k\leq2n$,
\[
{T_{k.k}}=\left(  {%
\begin{array}
[c]{ccccc}%
{{K_{0,0}^{(k)}}} & {{K_{0,1}^{(k)}}} &  &  & \\
& {{K_{1,1}^{(k)}}} & {{K_{1,2}^{(k)}}} &  & \\
&  & \ddots & \ddots & \\
&  &  & {{K_{n-1,n-1}^{(k)}}} & {{K_{n-1,n}^{(k)}}}\\
&  &  &  & {{K_{n,n}^{(k)}}}%
\end{array}
}\right)  ,
\]
for $0\leq i\leq n-1$,
\[
{K_{i,i+1}^{(k)}}={\left(  {%
\begin{array}
[c]{cccc}%
{(N-k-i)\gamma q} &  &  & \\
& {(N-k-i-1)\gamma q} &  & \\
&  & \ddots & \\
&  &  & {(N-k-n+1)\gamma q}\\
&  &  & 0
\end{array}
}\right)  _{(n+1-i)\times(n-i)}},
\]
\[
{K_{i,i}^{(k)}}={\left(  {%
\begin{array}
[c]{ccccc}%
{{c_{k,i,0}}} & {(N-k-i)\theta} &  &  & \\
\mu & {{c_{k,i,1}}} & {(N-k-i-1)\theta} &  & \\
& \ddots & \ddots & \ddots & \\
&  & {(n-i-1)\mu} & {{c_{k,i,n-i-1}}} & {(N-k-n+1)\theta}\\
&  &  & {(n-i)\mu} & {c_{k,i,n-i}}%
\end{array}
}\right)  _{(n+1-i)\times(n+1-i)}},
\]%
\[
{c_{k,i,m}}=-\left[  {(N-k-i-m)(\theta+\gamma)+m\mu}\right]  ,{0}\leq m\leq
n-i-1,
\]%
\[
{c_{k,i,n-i}}=-\left[  {(N-k-n)\gamma p+(n-i)\mu}\right]  ,
\]%
\[
{K_{n,n}^{(k)}} = - (N - k - n)\gamma p .
\]

Note that the block-generated time $W_{B}$ of any transaction package at the PBFT-based blockchain
system is a time interval from the arrival epoch of the new
transaction package to the time when it is determined as a block finally. Here, there is a reason why a transaction package becomes a block. That is, the transaction package obtains enough approval votes, so this transaction package or block can peg to the blockchain.

Let $\left(  {{\alpha_{0}},\alpha} \right)  $ denote the initial probability
distribution of the Markov process $\Psi$ with absorption state ${\Delta_{1}}$ at time $t=0$
for ${\alpha_{0}} = 0$. Then the vector $\alpha= \left(  {1,0, \ldots, 0} \right)
$ shows that the Markov process $T$ is at the state $\left(  {0,0,0}
\right)  $ at time 0. The following theorem provides an expression for the probability distribution of
the block-generated time $W_{B}$ by means of the first passage times and the
phase-type distributions of finite sizes.

\begin{The} \label{The-1}
If the initial probability distribution of the Markov process $\Psi$ with
absorption state ${\Delta_{1}}$ is $\left(  {{\alpha_{0}},\alpha} \right)  $ for $\alpha_0=0$,
then the probability distribution of the block-generated time $W_{B}$ is of
phase-type with an irreducible matrix representation $\left(  {\alpha,T}
\right)  $ of finite sizes, and
\[
{F_{{W_{B}}}}(t) = P\left\{  {{W_{B}} \le t} \right\}  = 1 -\alpha\exp\left\{
{Tt} \right\}  e,\quad t \ge0.
\]
Also, the average block-generated time is given by
\[
E\left[  {{W_{B}}} \right]  = - \alpha{T^{ - 1}}e,
\]
where $T^{ - 1}$ is the inverse of the matrix $T$ of finite sizes.
\end{The}

\textbf{Proof.} For $k \in\left\{  {0,1,2, \ldots,2n} \right\}  ,i \in\left\{
{0,1,2, \ldots,n} \right\}  ,j \in\left\{  {0,1,2, \ldots,n} \right\}  ,0 \le
i + j \le n$, we write
\[
{q_{k,i,j}}(t) = P\left\{  {N(t) = k,{J_{1}}(t) = i,{J_{2}}(t) = j} \right\}
,
\]
which is the state probability that the QBD process $\Psi$ with absorption
state ${\Delta_{1}}$ is at state $\left(  {k,i,j} \right)  $ at time $t \ge0$
before absorbed to state ${\Delta_{1}}$,
\[
{q_{k}}(t) = \left\{  {{q_{k,0,0}}(t), \ldots,{q_{k,0,n}}(t);{q_{k,1,0}}(t),
\ldots,{q_{k,1,n - 1}}(t); \ldots;{q_{k,n,0}}(t)} \right\}  ,
\]
and
\[
q(t) = \left\{  {{q_{0}}(t),{q_{1}}(t),{q_{2}}(t), \ldots,{q_{2n-1}}(t),{q_{2n}}(t)}
\right\}  .
\]
Using the Chapman-Kolmogorov forward differential equation, we can obtain
\begin{equation}
\frac{d}{{dt}}q(t) = q(t)T, \label{eq-2}%
\end{equation}
with the initial condition
\begin{equation}
q(0) = \alpha. \label{eq-3}%
\end{equation}
It follows from (\ref{eq-2}) and (\ref{eq-3}) together with $\alpha_0=0$ that
\begin{equation}
q(t) = \alpha\exp\{T t\}. \label{eq-4}%
\end{equation}
Thus we obtain
\[
P\left\{  {{W_{B}} > t} \right\}  = q(t)e = \alpha\exp\{T t\}e.
\]
This gives
\begin{align*}
F_{W_{B}}(t)  &  =P\left\{  W_{B} \leq t\right\}  =1-P\left\{  W_{B}>t\right\}
\\
&  =1-\alpha\exp\{T t\} e, \quad t \geq0
\end{align*}

In what follows, we compute the average block-generated time $E\left[  {{W_{B}%
}} \right]  $. Let $f(s)$ be the Laplace-Stieltjes transform of the
distribution function $F_{W_{B}}(t)$, then
\[
f(s) = \int_{0}^{\infty}{e^{ - st}}\textrm{d}F_{W_{B}}(t) = 1+ \alpha{(sI - T)^{ - 1}%
}{T^{0}}, \text{for } s \ge0,
\]
where $I$ denotes an identity matrix of finite size. Hence we obtain that
\[
E\left[  {{W_{B}}} \right]  = - \frac{\mathrm{d}}{{\mathrm{d}s}}f{(s)_{|s = 0}} = \alpha{\left[
{{{(sI - T)}^{ - 2}}} \right]  _{|s = 0}}{T^{0}} = - \alpha{T^{ - 1}}e
\]
by using ${T^{0} } + T\mathbf{{e}} = 0$ and ${T^{ - 1}}T\mathbf{{e}} =
\mathbf{{e}}$. This completes the proof. $\square$

To compute the inverse matrix $T^{ - 1}$ of finite size, we need to use the RG-factorizations of the Markov process $T$. To this end, we write
\[
{T^{-1}}=\left(
\begin{array}
[c]{ccccc}%
{{J_{0,0}}} & {{J_{0,1}}} & {{J_{0,2}}} & \cdots & {{J_{0,2n}}}\\
& {{J_{1,1}}} & {{J_{1,2}}} & \cdots & {{J_{1,2n}}}\\
&  & {{J_{2,2}}} & \cdots & {{J_{2,2n}}}\\
&  &  & \ddots & \vdots\\
&  &  &  & {{J_{2n,2n}}}%
\end{array}
\right)  ,
\]
By using $T^{-1}T=I$, we can obtain that for $k=0,1,\ldots,2n$,
\[
{J_{k,k}}=T_{k,k}^{-1},
\]
and for $k=0,1,\ldots,2n-1$, $j=1,2,\ldots,2n-k$,
\[
{J_{k,k+j}}={\left(  {-1}\right)  ^{j}}T_{k,k}^{-1}{T_{k,k+1}}T_{k+1,k+1}%
^{-1}{T_{k,k+2}}\cdots{T_{k+j-1,k+j}}T_{k+j,k+j}^{-1}.
\]
Therefore, before computing $T^{ - 1}$, it is a key to compute the inverse
matrices of diagonal block elements $T_{k,k}$ for $k = 0,1, \ldots,2n$.
Note that $K_{i,i}^{(k)}\neq\mathbf{{0}}$, then the upper triangular matrix
$T_{k,k}$ is invertible, and there exists a unique inverse matrix. Thus, we write
\[
T_{k,k}^{-1}=\left(  {%
\begin{array}
[c]{ccccc}%
{X_{0,0}^{(k)}} & {X_{0,1}^{(k)}} & {X_{0,2}^{(k)}} & \cdots & {X_{0,n}^{(k)}%
}\\
& {X_{1,1}^{(k)}} & {X_{1,2}^{(k)}} & \cdots & {X_{1,n}^{(k)}}\\
&  & {X_{2,2}^{(k)}} & \cdots & {X_{2,n}^{(k)}}\\
&  &  & \ddots & \vdots\\
&  &  &  & {X_{n,n}^{(k)}}%
\end{array}
}\right)  ,\quad0\leq k\leq2n.
\]
By using $T_{k,k}T_{k,k}^{-1}=I$, we can obtain that for $i=0,1,\ldots,n$,
\[
X_{i,i}^{(k)}={\left(  {K_{i,i}^{(k)}}\right)  ^{-1}},
\]
and for $i=0,1,\ldots,n-1$, $j=1,2,\ldots,n-i$,
\[
X_{i,i+j}^{(k)}={\left(  {-1}\right)  ^{j}}{\left(  {K_{i,i}^{(k)}}\right)
^{-1}}K_{i,i+1}^{(k)}{\left(  {K_{i+1,i+1}^{(k)}}\right)  ^{-1}}%
K_{i+1,i+2}^{(k)}\cdots K_{i+j-1,i+j}^{(k)}{\left(  {K_{i+j,i+j}^{(k)}%
}\right)  ^{-1}}.
\]
Therefore, before computing $T_{k,k}^{-1},0\leq k\leq2n$, it is a key to deal with
the inverse matrices of diagonal block elements $K_{i,i}^{(k)}$ with the order
$n+1-i$ for $i=0,1,\ldots,n$. Since $K_{i,i}^{(k)}$ is a birth and death process with finite states, we can use RG-factorizations to compute its inverse matrix. Readers may refer to Chapter 1 of Li \cite{Li:2010} for more details.

\subsection{The orphan-block-generated time distribution}

\label{subsec:orphan-generated time} 
\begin{figure}[pbth]
\centering        \includegraphics[width=12cm]{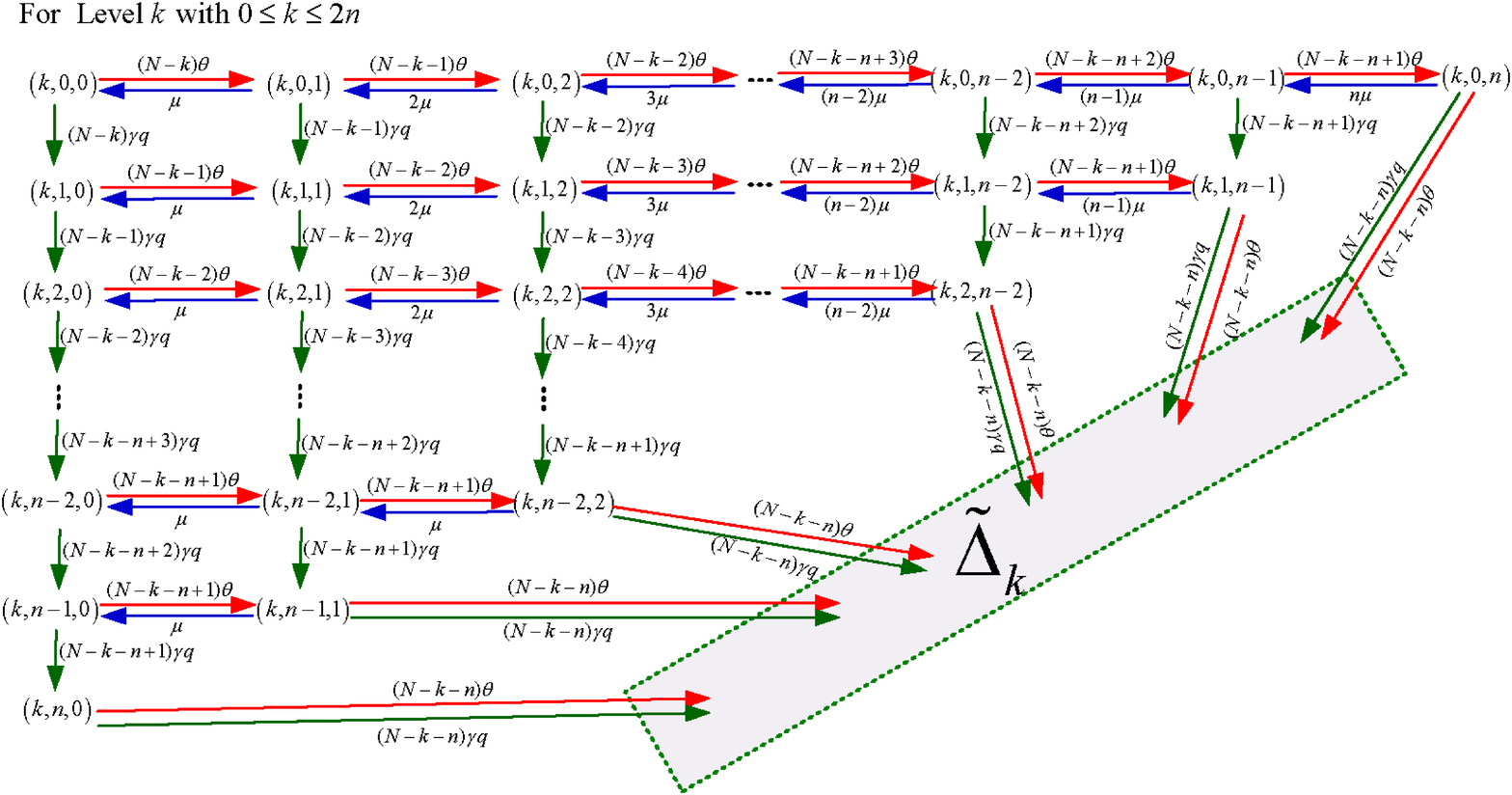}  \caption{The state transition relations of the Markov process $\Theta$: Part one.}%
\label{figure:Fig-5}%
\end{figure}

\begin{figure}[pbth]
\centering        \includegraphics[width=10cm]{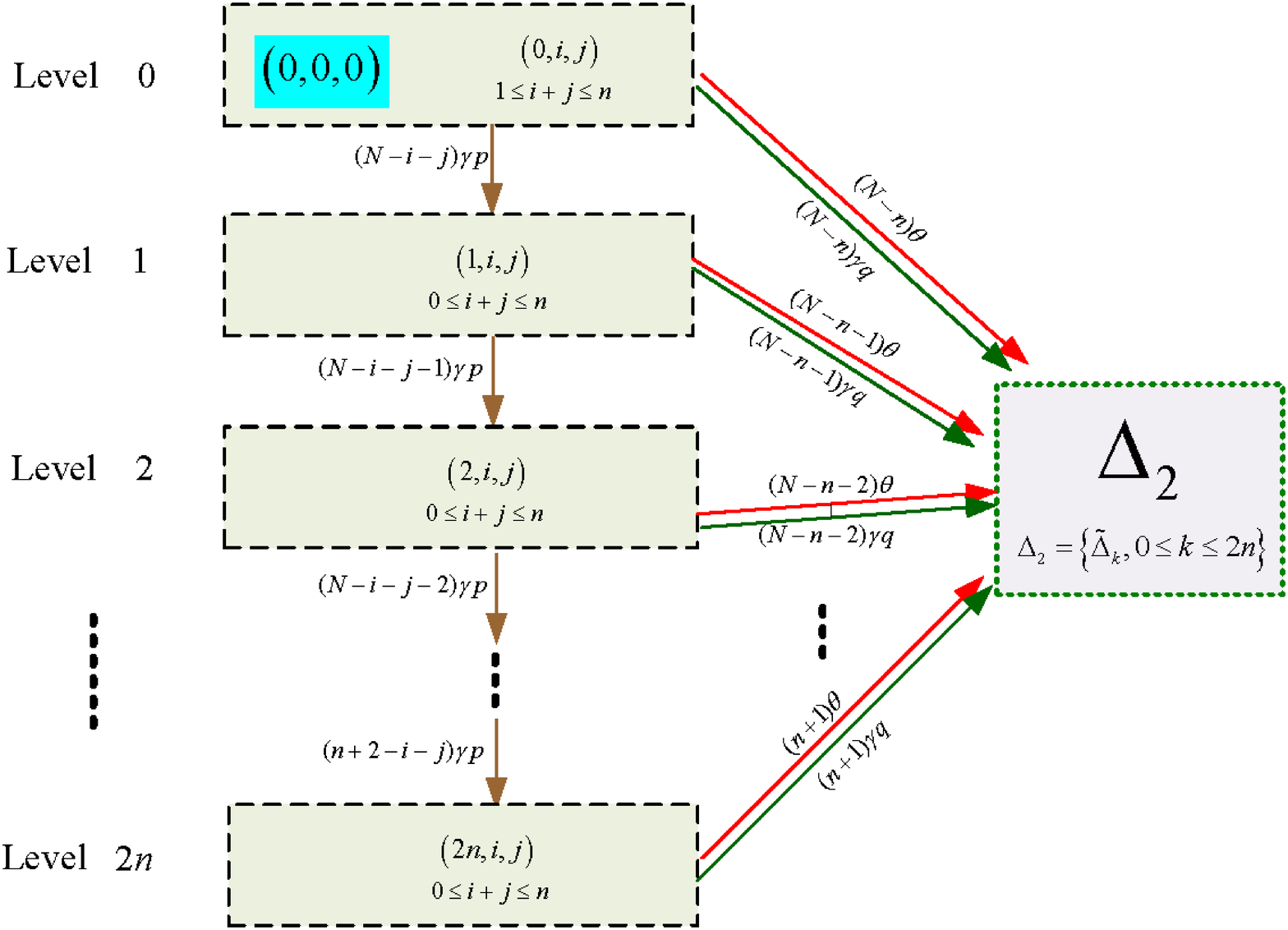}  \caption{The state transition relations of the Markov process $\Theta$: Part Two.}%
\label{figure:Fig-6}%
\end{figure}

To analyze the orphan-block-generated time
distribution, we set up a new Markov process with an absorption state
${\Delta_{2}}$, where, ${\Delta _2} = \left\{ {{{\tilde \Delta }_k},0 \le k \le 2n} \right\}$. To do this, all the states in the set
\[
\left\{  {(k,i,j):0 \le k \le2n,0 \le i,j \le n + 1,i + j = n + 1} \right\}
\]
are regarded as an absorption state ${\Delta_{2}}$. Then the Markov process
$\left\{  {\left(  {N(t),M(t),K(t)} \right)  :t \ge0} \right\}  $ operates on
a new state space with the absorption state ${\Delta_{2}}$ as follows:
\[
\left\{  {{\Delta_{2}}} \right\}  \cup\left\{  {(k,i,j):0 \le k \le 2n,0 \le i+j \le n   } \right\}  .
\]
In this case, the state transition relations of the Markov process $\left\{  {\left(
{N(t),M(t),K(t)} \right)  :t \ge0} \right\}  $ with the absorption state
${\Delta_{2}}$ are depicted in Figures \ref{figure:Fig-5} and
\ref{figure:Fig-6}. Also, its infinitesimal generator is given by


\[
\Theta=\left(
\begin{array}
[c]{cc}%
0 & 0\\
{{S^{0}}} & S
\end{array}
\right)  ,
\]
where ${S^{0}}+S\mathbf{{e}}=0$. In order to increase the readability of this paper, the non-zero matrix elements of $\Theta$ can be found in the Appendix A.

The orphan-block-generated time $W_{O}$ of any transaction package at the PBFT-based blockchain
system is a time interval from the arrival epoch of a new
transaction package to the time when it is determined as an orphan block finally. Here, there is a reason why a transaction package becomes an orphan block. That is, the transaction package cannot obtain enough approval votes, so this transaction package or orphan block needs to roll back to the transaction pool.

Let $\left(  {{\omega_{0}},\omega} \right)  $ denote the initial probability
distribution of the Markov process $\Theta$ with absorption state ${\Delta_{2}}$ at time $t=0$
for ${\omega_{0}} = 0$. Then the vector $\omega= \left(  {1,0, \ldots, 0} \right)
$ shows that the Markov process $S$ is at the state $\left(  {0,0,0}
\right)  $ at time 0.

The following theorem provides an expression for the probability distribution of
the orphan-block-generated time $W_{O}$ by means of the first passage times and the
phase-type distributions of finite sizes. The proof is omitted here to save space time since it is similar to that given in Theorem \ref{The-1}.

\begin{The}
If the initial probability distribution of the Markov process $\Theta$ with
absorption state ${\Delta_{2}}$ is $\left(  {{\omega_{0}},\omega} \right)  $ for $\omega_0=0$,
then the probability distribution of the orphan-block-generated time $W_{O}$ is of
phase-type with an irreducible matrix representation $\left(  {\omega,S}
\right)  $ of finite sizes, and
\[
{F_{{W_{O}}}}(t) = P\left\{  {{W_{O}} \le t} \right\}  = 1 -\omega\exp\left\{
{St} \right\}  e,\quad t \ge0.
\]
Also, the average orphan-block-generated time is given by
\[
E\left[  {{W_{O}}} \right]  = - \omega{S^{ - 1}}e.
\]
\end{The}

\section{Queueing Analysis for the PBFT-based Blockchain System}\label{sec:blockchain system}

In this section, we set up an $\mathrm{{M}} \oplus\mathrm{{P}}{\mathrm{{H}%
}^{b}}/\mathrm{{P}}{\mathrm{{H}}^{b}}/1$ queue to study the PBFT-based blockchain
system with repairable voting nodes. Based on this, we first give the
stationary probability vector of the queueing model for the PBFT-based blockchain system, and then provide performance analysis for the PBFT-based blockchain system
with repairable voting nodes.

\subsection{An $\mathrm{{M}} \oplus\mathrm{{P}}{\mathrm{{H}}^{b}}/\mathrm{{P}}{\mathrm{{H}}^{b}}/1$ queue}

From the block- and orphan-block-generated processes given in Section \ref{sec:distribution}, we set up an $\mathrm{{M}}
\oplus\mathrm{{P}}{\mathrm{{H}}^{b}}/\mathrm{{P}}{\mathrm{{H}}^{b}}/1$ queue
to provide performance analysis for the PBFT-based blockchain systems with
repairable voting nodes. To do this, the $\mathrm{{M}} \oplus\mathrm{{P}}{\mathrm{{H}}^{b}%
}/\mathrm{{P}}{\mathrm{{H}}^{b}}/1$ queue is described as follows:

\textbf{(1) Transaction arrivals at the transaction pool:} In the PBFT-based
blockchain system with repairable voting nodes, the arrival process of transactions contains two parts:

\textbf{(a)} The external transactions arrive at the transaction pool: We
assume that the external transactions arrive at the transaction pool according
to a Poisson process with an arrival rate $\lambda>0$.

\begin{figure}[pth]
\centering           \includegraphics[width=10cm]{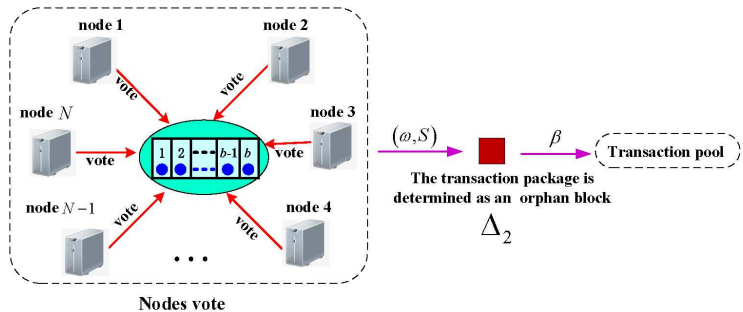}  \caption{The
process that the orphan block is rolled back to the transaction pool.}%
\label{figure:Fig-7}%
\end{figure}

\textbf{(b)} The orphan blocks are rolled back to the transaction pool: We subdivide the
rollback of every orphan block into two stages: The first stage is to determine the transaction package as an orphan block by the voting nodes; The
second stage is to roll back the orphan block to the transaction pool through
the network propagation. As seen from Subsection
\ref{subsec:orphan-generated time}, the orphan-block-generated time
$W_{O}$ follows a PH distribution with irreducible matrix representation $\left(
{\omega,S} \right)  $ of finite sizes, where the size of the orphan
block is $b$. In addition, we assume that the rollback time of the orphan
block is exponential with the rollback rate $\beta$. See Figure \ref{figure:Fig-7} for more details.

Combining Assumptions \textbf{(a)} with \textbf{(b)}, it is seen that the total
transaction arrivals at the PBFT-based blockchain system with repairable voting nodes are two processes: One is of phase type with irreducible matrix
representation $\left(  {\omega,S} \right)  $; while another is Poisson with arrival
rate $\lambda$.

\begin{figure}[pth]
\centering           \includegraphics[width=10cm]{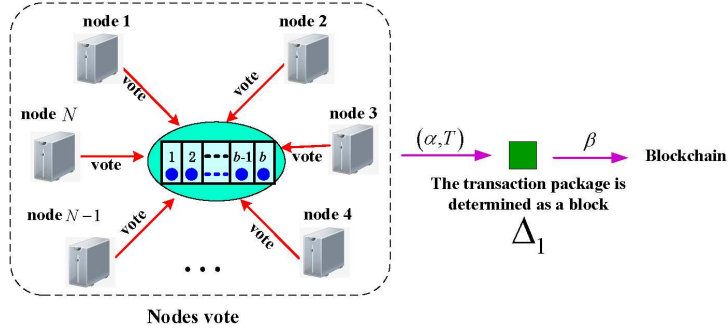}  \caption{Two different stages that the block enters the blockchain.}%
\label{figure:Fig-8}%
\end{figure}

\textbf{(2) The service times:} We regard the block-generated and block-pegged process as a two-stage
service process in the PBFT-based blockchain system with repairable voting nodes, that is, the service process contains two stages:
The first stage is that the PBFT-based blockchain system with repairable voting
nodes randomly selects $b$ transactions from the transaction pool with equal
probability to form a transaction package, and then this transaction package can be successfully
determined as a block by the voting nodes; the second stage is that this
block is pegged on the blockchain through the network propagation. Referring to Subsection \ref{subsec:block-generated time}, the
block-generated time $W_{B}$ of every transaction package at the PBFT-based blockchain
system follows a PH distribution with irreducible matrix representation $\left(  {\alpha,T} \right)  $ of finite sizes, where the size of the
orphan block is $b$. Also, we assume that the block-pegged time of the block
in the network follows an exponential distribution with the block-pegged rate $\beta$. See Figure \ref{figure:Fig-8} for more details.

\textbf{(3) Independence: }We assume that all random variables defined above
are independent of each other.

\begin{figure}[phtb]
\centering           \includegraphics[width=8cm]{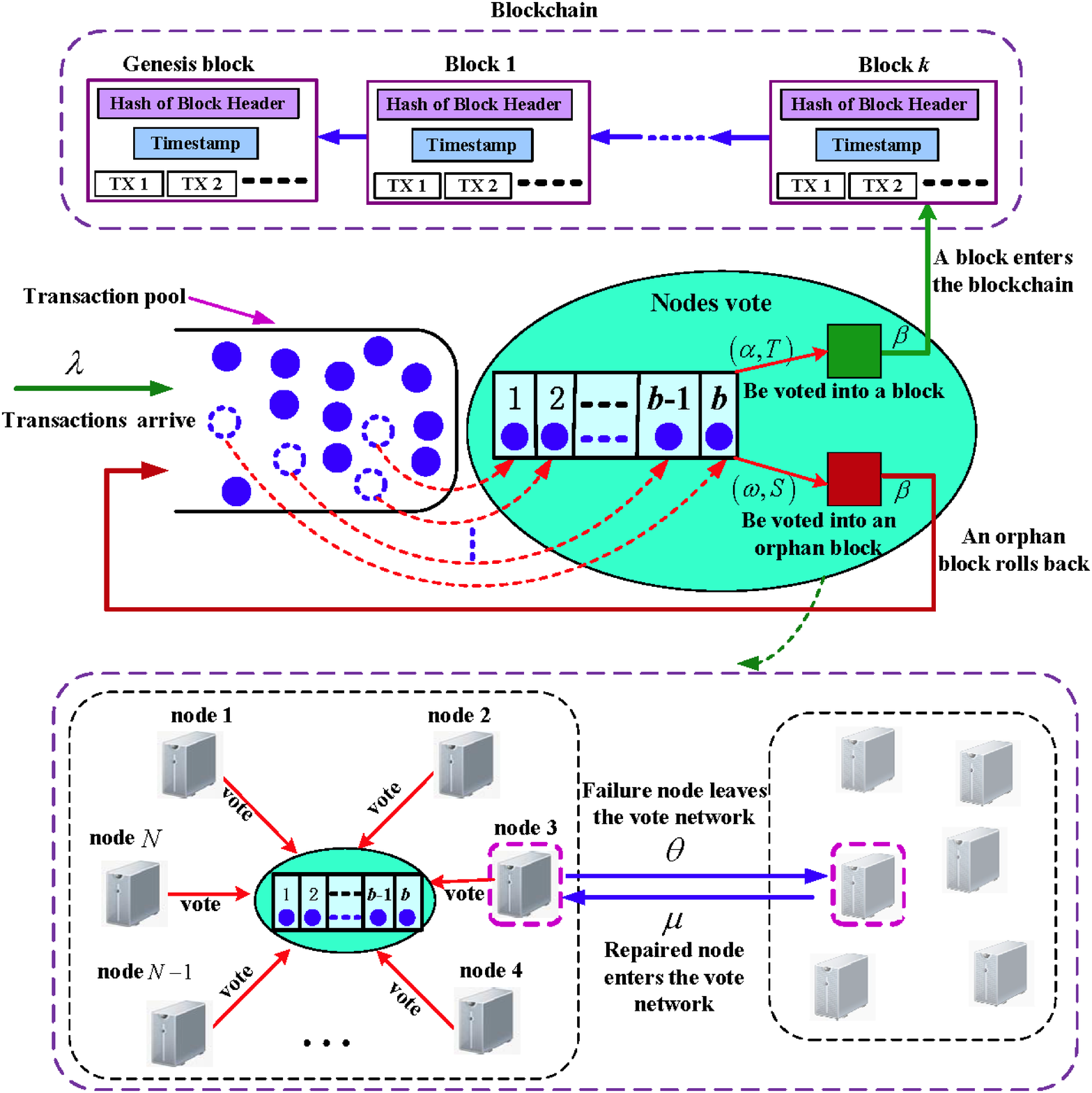}  \caption{The $\mathrm{{M}} \oplus\mathrm{{P}}{\mathrm{{H}}^{b}}/\mathrm{{P}}{\mathrm{{H}%
}^{b}}/1$ queue.}%
\label{figure:Fig-9}%
\end{figure}

From the above model assumptions, it is easy to see that the PBFT-based blockchain
system with repairable nodes can be described as an $\mathrm{{M}} \oplus
\mathrm{{P}}{\mathrm{{H}}^{b}}/\mathrm{{P}}{\mathrm{{H}}^{b}}/1$ queue, which is depicted in Figure \ref{figure:Fig-9}. To compute easily, we need to express the service time. Note that a block is pegged to the blockchain through two-stage processes: a PH distribution and
an exponential distribution, the total time of these two is written as a Markov process whose infinitesimal generator is given by:
\[
\left(  {%
\begin{array}
[c]{ccc}%
T & {{T^{0}}} & 0\\
0 & {-\beta} & \beta\\
0 & 0 & 0
\end{array}
}\right)  ,
\]
where the final state is an absorbing state.
Let
\[
\tilde{T}=\left(  {%
\begin{array}
[c]{cc}%
T & {{T^{0}}}\\
0 & {-\beta}%
\end{array}
}\right)  ,
\]
\[
{\tilde{T}^{0}}=\left(  {%
\begin{array}
[c]{c}%
0\\
\beta
\end{array}
}\right)  ,
\]
\[
\tilde{\alpha}=\left(  {\alpha,0}\right)  .
\]
It is clear that the time length that the block is generated and pegged to the blockchain follows a
continuous-time PH distribution with irreducible matrix representation
$\left(  {\tilde{\alpha},\tilde{T}}\right)  $.

Similarly, the time length that the orphan block is generated and rolled back to the transaction pool follows a continuous-time
PH distribution with irreducible matrix representation $\left(
{\tilde{\omega},\tilde{S}}\right)  $, where,
\[
\tilde{S}=\left(  {%
\begin{array}
[c]{cc}%
S & {{S^{0}}}\\
0 & {-\beta}%
\end{array}
}\right)  ,
\]
\[
\tilde{\omega}=\left(  {\omega,0}\right)  .
\]

\subsection{Analysis of the $\mathrm{{M}} \oplus\mathrm{{P}}{\mathrm{{H}}^{b}%
}/\mathrm{{P}}{\mathrm{{H}}^{b}}/1$ queue}

Let $I(t)$ be the number of transactions in the transactions pool at time $t$. Then
\[I(t) \in \left\{ {0,1,2, \ldots ,b - 1,b,b + 1,b + 2, \ldots } \right\}.\]
We denote by $C(t)$ and $D(t)$ the phases of the block-generated time and the orphan-block-generated time at time $t$, respectively. It is clear that
$\left\{ {\left( {I(t),C(t),D(t)} \right):t \ge 0} \right\}$ is a continuous-time Markov Process.

\begin{figure}[pbth]
\centering           \includegraphics[width=10cm]{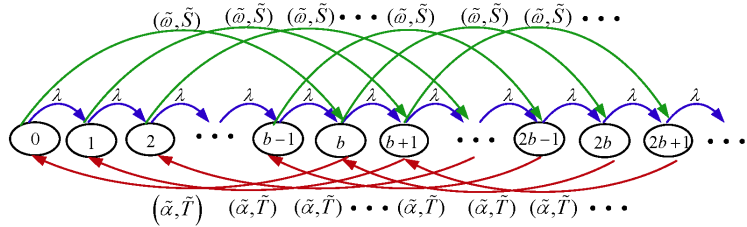}  \caption{The state transition relations of the Markov process $\left\{ {\left( {I(t),C(t),D(t)} \right):t \ge 0} \right\}$.}%
\label{figure:Fig-10}%
\end{figure}

Based on Figure \ref{figure:Fig-10}, the infinitesimal generator $\Upsilon$ of
the Markov process $\left\{ {\left( {I(t),C(t),D(t)} \right):t \ge 0} \right\}$ is given by%

\[
\Upsilon=\left(  {%
\begin{array}
[c]{ccccc}%
{B_{1}^{(0)}} & {A_{0}^{(0)}} &  &  & \\
{A_{2}^{(1)}} & {{A_{1}}} & {{A_{0}}} &  & \\
& {{A_{2}}} & {{A_{1}}} & {{A_{0}}} & \\
&  & \ddots & \ddots & \ddots
\end{array}
}\right)  ,
\]
where
\[
B_{1}^{(0)}=\left(  {%
\begin{array}
[c]{ccccc}%
{\tilde{S}-\lambda I} & {\lambda I} &  &  & \\
& {\tilde{S}-\lambda I} & {\lambda I} &  & \\
&  & \ddots & \ddots & \\
&  &  & {\tilde{S}-\lambda I} & {\lambda I}\\
&  &  &  & {\tilde{S}-\lambda I}%
\end{array}
}\right)  ,
\]%

\[
A_{0}^{(0)}=\left(  {%
\begin{array}
[c]{cccc}%
{({{\tilde{S}}^{0}}\tilde{\omega})\otimes\tilde{\alpha}} &  &  & \\
& {({{\tilde{S}}^{0}}\tilde{\omega})\otimes\tilde{\alpha}} &  & \\
&  & \ddots & \\
{\lambda(I\otimes\tilde{\alpha})} &  &  & {({{\tilde{S}}^{0}}\tilde{\omega
})\otimes\tilde{\alpha}}%
\end{array}
}\right)  ,
\]%

\[
A_{2}^{(1)}=\left(  {%
\begin{array}
[c]{cccc}%
{I\otimes{{\tilde{T}}^{0}}} &  &  & \\
& {I\otimes{{\tilde{T}}^{0}}} &  & \\
&  & \ddots & \\
&  &  & {I\otimes{{\tilde{T}}^{0}}}%
\end{array}
}\right)  ,
\]%

\[
{A_{2}}=\left(  {%
\begin{array}
[c]{cccc}%
{I\otimes\left(  {{{\tilde{T}}^{0}}\tilde{\alpha}}\right)  } &  &  & \\
& {I\otimes\left(  {{{\tilde{T}}^{0}}\tilde{\alpha}}\right)  } &  & \\
&  & \ddots & \\
&  &  & {I\otimes\left(  {{{\tilde{T}}^{0}}\tilde{\alpha}}\right)  }%
\end{array}
}\right)  ,
\]%

\[
{A_{1}}=\left(  {%
\begin{array}
[c]{ccccc}%
{\tilde{S}\oplus\tilde{T}-\lambda I} & {\lambda I} &  &  & \\
& {\tilde{S}\oplus\tilde{T}-\lambda I} & {\lambda I} &  & \\
&  & \ddots & \ddots & \\
&  &  & {\tilde{S}\oplus\tilde{T}-\lambda I} & {\lambda I}\\
&  &  &  & {\tilde{S}\oplus\tilde{T}-\lambda I}%
\end{array}
}\right)  ,
\]%

\[
{A_{0}}=\left(  {%
\begin{array}
[c]{cccc}%
{({{\tilde{S}}^{0}}\tilde{\omega})\otimes I} &  &  & \\
& {({{\tilde{S}}^{0}}\tilde{\omega})\otimes I} &  & \\
&  & \ddots & \\
{\lambda I} &  &  & {({{\tilde{S}}^{0}}\tilde{\omega})\otimes I}%
\end{array}
}\right)  ,
\]
and $I$ is the identity matrix of appropriate dimensions.

Obviously, the continuous-time Markov process $\Upsilon$ is a
level-independent QBD process. Thus, we can apply the matrix-geometric
solution given in Neuts \cite{Neuts:1981} to analyze the QBD process $\Upsilon$. Based on this, we can further analyze the PBFT-based blockchain system
with repairable voting nodes.

\begin{The}
The level-independent QBD $\Upsilon$ is positive recurrent if and only if
\[
\lambda+b\delta_{1}{\tilde S}^{0}<b\delta_{2}{\tilde T}^{0}  ,
\]
where ${\delta_{1}}$ and ${\delta_{2}}$ are the stationary probability vectors of the two Markov processes $\tilde T + {\tilde T^{0}}\tilde\alpha$ and $\tilde S + {\tilde
S^{0}}\tilde\omega$, respectively.
\end{The}

\textbf{Proof.} Let $W = (\tilde S + {\tilde S^{0}}\tilde\omega) \oplus(\tilde
T + {\tilde T^{0}}\tilde\alpha)$. Since ${\delta_{1}}$ satisfies ${\delta_{1}}(\tilde T + {\tilde T^{0}}\tilde\alpha) = 0$, ${\delta_{1}}e =
1$; and ${\delta_{2}}$ satisfies ${\delta_{2}}(\tilde S + {\tilde
S^{0}}\tilde\omega) = 0$, ${\delta_{2}}e = 1$. It is easy to see that $({\delta_{1}} \otimes
{\delta_{2}})\left[  {(\tilde S + {{\tilde S}^{0}}\tilde\omega) \oplus(\tilde
T + {{\tilde T}^{0}}\tilde\alpha)} \right]  = 0$.

For the continuous-time QBD process $\Upsilon$, we use the mean-drift method
to provide its stability condition. Readers may refer to Chapter 1 of Neuts
\cite{Neuts:1981} or Chapter 3 of Li \cite{Li:2010} for more details. We write
\[
A={A_{2}}+{A_{1}}+{A_{0}}=\left(  {%
\begin{array}
[c]{ccccc}%
{W-\lambda I} & {\lambda I} &  &  & \\
& {W-\lambda I} & {\lambda I} &  & \\
&  & \ddots & \ddots & \\
&  &  & {W-\lambda I} & {\lambda I}\\
{\lambda I} &  &  &  & {W-\lambda I}%
\end{array}
}\right)  ,
\]
Clearly, the Markov process $A$ is irreducible, aperiodic and positive
recurrent. Let $\upsilon  = \left( {{\upsilon _1},{\upsilon _2}, \ldots ,{\upsilon _b}} \right)$ be the stationary probability vector of Markov
process $A$, where ${\upsilon_{k}}={\gamma_{k}}({\delta_{1}}\otimes{\delta_{2}%
}),k=1,2,\ldots,b$. Then from the system of linear equations: $\upsilon A=0$
and $\upsilon e=1$, we easily get that ${\gamma_{1}}={\gamma_{2}}=\cdots
{\gamma_{b}}=1/b$, thus, we obtain ${\upsilon_{k}}=({\delta_{1}%
}\otimes{\delta_{2}})/b,k=1,2,\ldots,b$.

Using the mean-drift method, it is easy to check that the QBD process $\Upsilon$
is positive recurrent if and only if $\upsilon{A_{0}}e < \upsilon{A_{2}}e$,
i.e.,
\[
\lambda+b\delta_{1}{\tilde S}^{0}<b\delta_{2}{\tilde T}^{0}  .
\]
This completes the proof. $\square$

When the QBD process $\Upsilon$ is positive recurrent, we write its stationary
probability vector as $\psi= \left(  {{\psi_{0}},{\psi_{1}},{\psi_{2}},
\ldots} \right)  $, where
\[
{\psi_{k}} = \left(  {{\tilde{\psi}_{kb}},{\tilde{\psi}_{kb + 1}}, \ldots,{\tilde{\psi}_{(k + 1)b -
1}}} \right)  ,k \ge0.
\]

Note that such a stationary probability vector $\psi$ in general has no explicit
expression, we can only develop its numerical solution. To this end,
by using the Chapter 3 in Neuts \cite{Neuts:1981}, we need to numerically
compute the rate matrix $R$, which is the minimal nonnegative solution to the
nonlinear matrix equation ${R^{2}}{A_{2}} + R{A_{1}} + {A_{0}} = 0$. Based on this, we give an iterative algorithm (see Chapter 3 of Neuts
\cite{Neuts:1981}) to numerically compute the rate matrix $R$ as follows:
\[
{R_{0}} = 0,
\]
\begin{equation}
{R_{n + 1}} = (R_{n}^{2}{A_{2}} + {A_{0}})\left(  { - A_{1}^{ - 1}} \right)
,n = 1,2,3, \ldots. \label{equa-5}
\end{equation}
For the matrix sequence $\left\{  {{R_{n}},n \ge0} \right\}  $, by means of Chapter 3 in Neuts \cite{Neuts:1981}, it is easy to see that as $n
\to\infty$, ${R_{n}} \uparrow R$. Thus, for any sufficiently small positive number
$\varepsilon$, there exists a positive integer $\mathbf{{n}}$ such that
$\left\|  {{R_{\mathbf{{n}} + 1}} - {R_{\mathbf{{n}}}}} \right\|  <
\varepsilon$. In this case, we take $R \approx{R_{\mathbf{{n}}}}$, which gives
an approximate solution to the nonlinear matrix equation ${R^{2}}{A_{2}} +
R{A_{1}} + {A_{0}} = 0$.

Using the rate matrix $R$, the
following theorem provides an expression for the stationary probability vector
$\psi$. This conclusion is directly derived from Theorem 1.2.1 of Chapter 1 of
Netus \cite{Neuts:1981}. Here, we restate it without a proof.

\begin{The}
If the QBD process $\Upsilon$ is positive recurrent, then its stationary
probability vector $\psi= \left(  {{\psi_{0}},{\psi_{1}},{\psi_{2}}, \ldots}
\right)  $ is given by
\begin{equation}
{\psi_{k}} = {\psi_{1}}{R^{k - 1}},k \ge1, \label{equa-6}
\end{equation}
where ${\psi_{0}}$ and ${\psi_{1}}$ are the unique solution to the following
system of linear equations:
\begin{equation}
\left\{
\begin{array}
[c]{l}%
{\psi_{0}}B_{1}^{(0)} + {\psi_{1}}A_{2}^{(1)} = 0,\\
{\psi_{0}}A_{0}^{(0)} + {\psi_{1}}({A_{1}} + R{A_{2}}) = 0,\\
{\psi_{0}}e + {\psi_{1}}{(I - R)^{ - 1}}e = 1.
\end{array}
\right.  \label{equa-7}
\end{equation}
\end{The}

\subsection{Performance measures of the PBFT system}

Based on the $\mathrm{{M}} \oplus\mathrm{{P}}{\mathrm{{H}}^{b}}/\mathrm{{P}%
}{\mathrm{{H}}^{b}}/1$ queue and the stationary probability vector $\psi$, we can
provide some key performance measures of the PBFT-based blockchain system with
repairable voting nodes as follows:

\textbf{(a1) The stationary probability of no transaction package in the
PBFT-based blockchain system} is given by
\[
{\eta_{1}} = {\psi_{0}}e.
\]

\textbf{(a2) The stationary probability of existing transaction package in the
PBFT-based blockchain system} is given by
\[
{\eta_{2}} = 1 - {\eta_{1}} = 1 - {\psi_{0}}e.
\]

\textbf{(b1) The stationary rate that a block is pegged on the blockchain
in the PBFT-based blockchain system} is given by
\[
{r_{1}} = \left(  {1 - {\psi_{0}}e} \right)  \frac{1}{{( - \tilde
\alpha{{\tilde T}^{ - 1}}e)}} = {\eta_{2}}\frac{1}{{( - \tilde\alpha{{\tilde
T}^{ - 1}}e)}}.
\]

\textbf{(b2) The stationary rate that an orphan block is rolled back to the
transaction pool} is given by
\[
{r_{2}} = \left(  {1 - {\psi_{0}}e} \right)  \frac{1}{{( - \tilde
\omega{{\tilde S}^{ - 1}}e)}} = {\eta_{2}}\frac{1}{{( - \tilde\omega{{\tilde
S}^{ - 1}}e)}}.
\]

\textbf{(c) The block throughput of the PBFT-based blockchain system} is given by
\[
\mathrm{{TH}(block)} = \left(  {1 - {\psi_{0}}e} \right)  \frac{1}{{( - \tilde
\alpha{{\tilde T}^{ - 1}}e)}} = {\eta_{2}}\frac{1}{{( - \tilde\alpha{{\tilde
T}^{ - 1}}e)}}.
\]

In this paper, we define the block throughput as the number of blocks per second, and transaction throughput as the number of transactions per second. The latter is a common and general definition of throughput and the focus of our attention.

In what follows, we provide an effective method to compute the transaction throughput of the PBFT-based
blockchain system with repairable voting nodes.

\begin{The}
The transaction throughput of the PBFT-based blockchain system is given by
\[
\mathrm{{TH}} = \left(  {1 - {\psi_{0}}e} \right)  \frac{b}{{( - \tilde
\alpha{{\tilde T}^{ - 1}}e)}} = {\eta_{2}}\frac{b}{{( - \tilde\alpha{{\tilde
T}^{ - 1}}e)}}.
\]
\end{The}

\textbf{Proof.} We only consider the transaction throughput $\mathrm{{TH}}$. From Theorem \ref{The-1} in Subsection \ref{subsec:block-generated time},
it is seen that the block-generated time $W_{B}$ of any transaction package
in the PBFT-based blockchain system follows a PH distribution with irreducible
matrix representation $\left(  {\tilde\alpha,\tilde T} \right)  $. This gives that the average time that the PBFT-based
blockchain system pegs a block with $b$ transactions on the blockchain is $( -
\tilde\alpha{\tilde T^{ - 1}}e)$. Therefore, the number of transactions dealt with
by the PBFT-based blockchain system per unit time is $ b/{(
- \tilde\alpha{{\tilde T}^{ - 1}}e)}$.

In addition, the stationary probability of the existing transaction package in the
PBFT-based blockchain system is given by ${\eta_{2}}$. Thus, the transaction
throughput of the PBFT-based blockchain system is the product of
the stationary probability of the existing transaction package in the PBFT-based
blockchain system and the number of transactions dealt with per
unit time in the PBFT-based blockchain system, thus, we obtain
\[
\mathrm{{TH}} = \left(  {1 - {\psi_{0}}e} \right)  \frac{b}{{( - \tilde
\alpha{{\tilde T}^{ - 1}}e)}} = {\eta_{2}}\frac{b}{{( - \tilde\alpha{{\tilde
T}^{ - 1}}e)}}.
\]
This completes the proof. $\square$

\section{Reliability Analysis of the PBFT-based Blockchain System} \label{sec:security and reliability}

In this section, we set up two new Markov processes to analyze the reliability of the PBFT-based blockchain system with repairable voting nodes. Here, our purpose focuses on such a condition: is considered to be the inability of the blockchain system to produce any blocks. In what follows, we consider two different cases.

\subsection{Unavailability due to failed nodes} \label{subsec:simple reliability}
In this subsection, we consider the case where the PBFT-based blockchain system becomes unavailable due to the number of failed nodes reaching $n + 1$. In this case, the PBFT-based blockchain system can no longer carry out the voting process such that no new blocks are generated.

\begin{figure}[pbth]
\centering           \includegraphics[width=12cm]{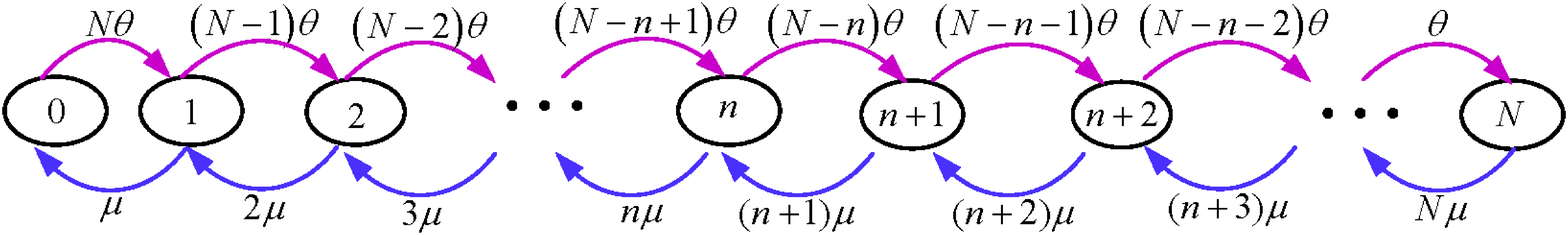}  \caption{The state transition relations of the birth-death process $\left\{ {K(t),t \ge 0} \right\}$.}%
\label{figure:Fig-11}%
\end{figure}

Note that every node may fail and need to be repaired, with the failure rate of $\theta$ and the repair rate of $\mu$. Let $K(t)$ represent the number of failed nodes in the PBFT-based blockchain system at time $t$, then $\left\{ {K(t),t \ge 0} \right\}$ is a continuous-time birth-death process, whose state space is $\Phi  = \left\{ {0,1,2, \ldots ,N} \right\}$, and its state transition relations are depicted in Figure \ref{figure:Fig-11}.

Based on Figure \ref{figure:Fig-11}, the infinitesimal generator of the birth-death process $\left\{ {K(t),t \ge 0} \right\}$ is given by
\[{Q_A} = \left( {\begin{array}{*{20}{c}}
{ - N\theta }&{N\theta }&{}&{}&{}&{}\\
\mu &{ - \left[ {\mu  + (N - 1)\theta } \right]}&{(N - 1)\theta }&{}&{}&{}\\
{}&{2\mu }&{ - \left[ {2\mu  + (N - 2)\theta } \right]}&{(N - 2)\theta }&{}&{}\\
{}&{}& \ddots & \ddots & \ddots &{}\\
{}&{}&{}&{(N - 1)\mu }&{ - \left[ {(N - 1)\mu  + \theta } \right]}&\theta \\
{}&{}&{}&{}&{N\mu }&{ - N\mu }
\end{array}} \right).\]

Obviously, the birth-death process $K(t)$ is aperiodic, irreducible, and positive recurrent. Let $\zeta  = \left( {{\zeta _0},{\zeta _1}, \ldots ,{\zeta _N}} \right)$ be the stationary probability vectors of the birth-death process $\left\{ {K(t),t \ge 0} \right\}$. Then we have the system of linear equations: $\zeta {Q_A} = 0$ and $\zeta e = 1$. Thus, we obtain
\[{\zeta _{k}} = \frac{{(N-k+1)\theta }}{{k\mu}}{\zeta _{k-1}},\quad 1 \le k \le N,\]
and
\[{\zeta _0} = \frac{1}{{1 + \sum\limits_{k = 1}^N {\frac{{N!}}{{(N - k)!k!}}{\rho ^k}} }},\]
where $\rho  = \theta /\mu $.

Let $A_{1}$ be the inherent stationary availability of the PBFT-based blockchain system with repairable voting nodes. Then
\[A_1 = {\zeta _0} + {\zeta _1} + {\zeta _2} +  \cdots  + {\zeta _n} = \frac{{1 + \sum\limits_{k = 1}^n {\frac{{N!}}{{(N - k)!k!}}{\rho ^k}} }}{{1 + \sum\limits_{k = 1}^N {\frac{{N!}}{{(N - k)!k!}}{\rho ^k}} }} .\]

\begin{figure}[pbth]
\centering           \includegraphics[width=8cm]{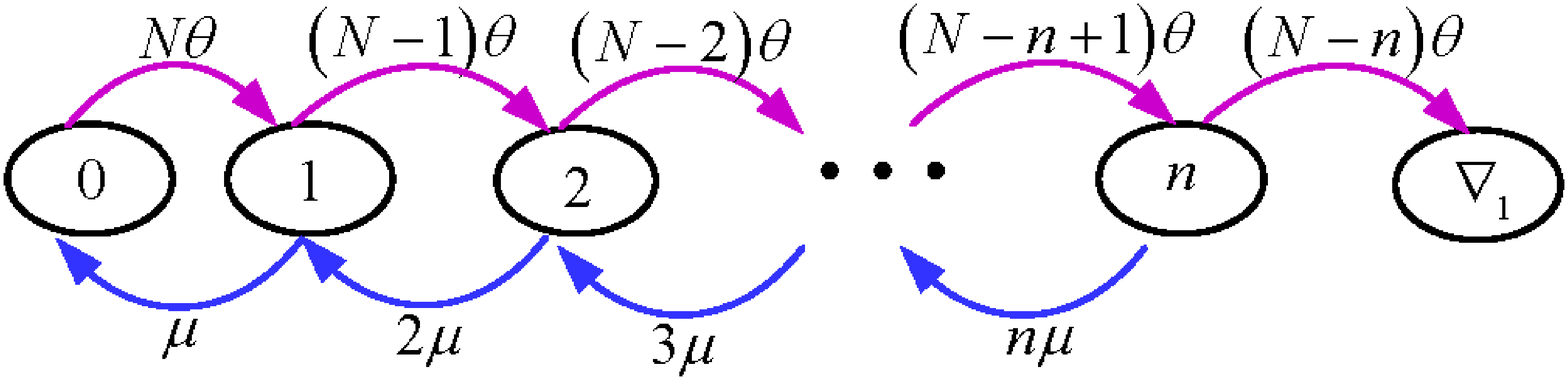}  \caption{The state transition relations of the Markov process $\left\{ {K(t),t \ge 0} \right\}$ with the absorption state $\nabla_1$.}%
\label{figure:Fig-12}%
\end{figure}

To analyze the inherent reliability $R_1(t)$ of the PBFT-based blockchain system with repairable voting nodes, we let all the states in the set $\left\{ {n + 1,n + 2, \ldots ,N} \right\}$ be absorption state $\nabla_1$. Then the Markov process $\left\{ {{K}(t),t \ge 0} \right\}$ operates on a new state space $\left\{ {0,1,2, \ldots ,n} \right\} \cup \left\{ \nabla_1  \right\}$, and its state transition relations are depicted in Figure \ref{figure:Fig-12}.

We write
\[{Q_{\nabla_1} } = \left( {\begin{array}{*{20}{c}}
{{T_\nabla }}&{T_\nabla ^0}\\
0&0
\end{array}} \right),\]
where,
\[{T_\nabla } = \left( {\begin{array}{*{20}{c}}
{ - N\theta }&{N\theta }&{}&{}&{}\\
\mu &{ - \left[ {\mu  + (N - 1)\theta } \right]}&{(N - 1)\theta }&{}&{}\\
{}& \ddots & \ddots & \ddots &{}\\
{}&{}&{(n - 1)\mu }&{ - \left[ {(n - 1)\mu  + (N - n + 1)\theta } \right]}&{(N - n + 1)\theta }\\
{}&{}&{}&{n\mu }&{ - \left[ {n\mu  + (N - n)\theta } \right]}
\end{array}} \right),\]
\[T_\nabla ^0 = \left( {\begin{array}{*{20}{c}}
0\\
 \vdots \\
0\\
{(N - n)\theta }
\end{array}} \right).\]

Let $(\varphi (0),{\varphi _0}(0))$ be the initial probability distribution of the Markov process $Q_{\nabla_1}$, where ${\varphi _0}(0) = 0$, and $\varphi (0) = \left( {1,0,0, \ldots ,0} \right)$, $\tau_1  = \inf \left\{ {t \ge 0:\tilde{K}(t) = n + 1,\tilde{K}(0) = 0} \right\}$, and $R_1(t) = P\left\{ {\tau_1  > t} \right\}$, then following theorem provides an expression for the inherent reliability $R_1(t)$ of the PBFT-based blockchain system.

\begin{The} \label{The-6}
If the initial probability distribution of the Markov process $Q_{\nabla_1}$ is $(\varphi (0),{\varphi _0}(0))$, then the inherent reliability $R_1(t)$ of the PBFT-based blockchain system is given by
\begin{equation}
R_1(t) =  \sum\limits_{k = 0}^n {{\varphi _k}(t)}  \label{eq-5}
\end{equation}
and $\varphi(t)=(\varphi _0(t),\varphi _1(t),\ldots,\varphi _n(t))$ satisfy the Chapman-Kolmogorov forward differential equation
\begin{equation}
\varphi '(t) = \varphi (t){T_\nabla } \label{eq-6}
\end{equation}
with $(\varphi (0),{\varphi _0}(0))=(1,0,0,\ldots ,0)$.
\end{The}

\textbf{Proof.} It is easy to see that $R_1(t) = \varphi (t)e$. Let
\[\varphi ^*(s) = \int_0^\infty  {{e^{ - st}}{\varphi }(t)dt,\quad s > 0, \quad i = 0,1,2, \ldots ,n}, \]
be the Laplace transform of ${\varphi }(t)$, then
\[\int_0^\infty  {{e^{ - st}}\varphi '(t)dt}  = \int_0^\infty  {{e^{ - st}}\varphi (t)dt}  \cdot {T_\nabla }, \quad s > 0,\]
thus, we obtain
\[{\varphi ^*}(s) = \int_0^\infty  {{e^{ - st}}\varphi (t)dt = } \varphi (0){(sI - {T_\nabla })^{ - 1}}, \quad s > 0.\]
Also, we have
\[{R_1^*}(s) = {\varphi ^*}(s)e = \varphi (0){(sI - {T_\nabla })^{ - 1}}e, \quad s > 0.\]
Therefore,
\[R_1(t) = P\left\{ {\tau_1  > t} \right\} = \sum\limits_{k = 0}^n {{\varphi _k}(t)}. \]
This completes the proof. $\square$

Based on Theorem \ref{The-6}, the average time before the first failure of the PBFT-based blockchain system is given by
\[{\rm{MTTFF_1 = }}\int_0^\infty  {R_1(t)dt = {R_1^*}(0) =  - \varphi (0)T_\nabla ^{ - 1}e} .\]

\subsection{Unavailability due to both failed nodes and disapproval votes } \label{subsec:complex reliability}
In this subsection, we consider the case where the PBFT-based blockchain system becomes unavailable due to the sum of failed nodes and disapproval votes reaching $n+1$. To this end, we use a three-dimension Markov process to analyze the reliability of the PBFT voting process with repairable voting nodes.

From Figures \ref{figure:Fig-1} and \ref{figure:Fig-2},  it is easy to see that the infinitesimal
generator of the Markov process $\left\{  {\left(  {N(t),M(t),K(t)} \right)  :t \ge0} \right\}$ is given by
\begin{equation}
Q = \left(  {%
\begin{array}
[c]{cccccc}%
{{Q_{0,0}}} & {{Q_{0,1}}} & {} & {} & {} & {}\\
{{Q_{1,0}}} & {{Q_{1,1}}} & {{Q_{1,2}}} & {} & {} & {}\\
{{Q_{2,0}}} & {} & {{Q_{2,2}}} & {{Q_{2,3}}} & {} & {}\\
\vdots & {} & {} & \ddots & \ddots & {}\\
{{Q_{2n,0}}} & {} & {} & {} & {{Q_{2n,2n}}} & {{Q_{2n,2n + 1}}}\\
{{Q_{2n + 1,0}}} & {} & {} & {} & {} & {{Q_{2n + 1,2n + 1}}}%
\end{array}
} \right)  . \label{eq-1}%
\end{equation}
Where the non-zero matrix elements of the infinitesimal generator $Q$ are given in the Appendix B.

Note that the Markov process $\left\{  {\left(  {N(t),M(t),K(t)} \right)  :t
\ge0} \right\}  $ is irreducible and it contains only finite states, thus it
is positive recurrent. By corresponding to state space $\Omega$, we define the
probabilities
\[
{\boldsymbol{\pi}  }=\left(  {{\boldsymbol{\pi}_{0}},{\boldsymbol{\pi}_{1}},{\boldsymbol{\pi}_{2}},\ldots,{\boldsymbol{\pi}_{2n}%
},{\boldsymbol{\pi}_{2n+1}}}\right)  ,
\]
where
\[
{{\boldsymbol{\pi}  }_{0}}=\left(  {{\pi_{0,0,0}},{\pi_{0,0,1}},\ldots
,{\pi_{0,0,n+1}};{\pi_{0,1,0}},{\pi_{0,1,1}},\ldots,{\pi_{0,1,n}};\ldots
;{\pi_{0,n,0}},{\pi_{0,n,1}};{\pi_{0,n+1,0}}}\right)  ,
\]%
\[
{{\boldsymbol{\pi}  }_{1}}=\left(  {{\pi_{1,0,0}},{\pi_{1,0,1}},\ldots
,{\pi_{1,0,n+1}};{\pi_{1,1,0}},{\pi_{1,1,1}},\ldots,{\pi_{1,1,n}};\ldots
;{\pi_{1,n,0}},{\pi_{1,n,1}};{\pi_{1,n+1,0}}}\right)  ,
\]%
\[
\vdots
\]%
\[
{{\boldsymbol{\pi}  }_{2n}}=\left(  {{\pi_{2n,0,0}},{\pi_{2n,0,1}},\ldots
,{\pi_{2n,0,n+1}};{\pi_{2n,1,0}},{\pi_{2n,1,1}},\ldots,{\pi_{2n,1,n}}%
;\ldots;{\pi_{2n,n,0}},{\pi_{2n,n,1}};{\pi_{2n,n+1,0}}}\right)  ,
\]%
\[
{{\boldsymbol{\pi}  }_{2n+1}}=\left(  {{\pi_{2n+1,0,0}},{\pi_{2n+1,0,1}%
},\ldots,{\pi_{2n+1,0,n}};{\pi_{2n+1,1,0}},{\pi_{2n+1,1,1}},\ldots
,{\pi_{2n+1,1,n-1}};\ldots;{\pi_{2n+1,n,0}}}\right)  .
\]

Since ${\boldsymbol{\pi}  }$ is the stationary probability vector of the Markov process $Q$, to express the stationary probability vector ${\boldsymbol{\pi}  }=\left(  {{\boldsymbol{\pi}_{0}},{\boldsymbol{\pi}_{1}},{\boldsymbol{\pi}_{2}},\ldots,{\boldsymbol{\pi}_{2n}%
},{\boldsymbol{\pi}_{2n+1}}}\right)  $, we
need to compute the inverse matrices for the matrices $Q_{k,k}$ in the
infinitesimal generator $Q$ given in equation (\ref{eq-1}) for $1 \le k \le2n
+ 1$. To do this, we observe that the structure of $Q_{k,k}$ is the same as that of
matrix $T_{i,i}$, the computation of the inverse matrix $Q_{k,k}^{ - 1}$
by RG-factorizations is similar to the inverse matrix $T_{i,i}^{ - 1}$. Thus,
we omit the details of computing the inverse matrix $Q_{k,k}^{ - 1}$. Readers may refer to Subsection \ref{subsec:block-generated time}.

Let ${R_{k}} = {Q_{k,k +1}}{( - {Q_{k+1,k+1}^{ - 1}})}$ for $ 0 \le k \le 2n $.
Next, we provide an expression for the stationary probability
vector ${\boldsymbol{\pi}  }$ using the following theorem.
\begin{The}
The stationary probability vector $\boldsymbol{\pi}  =\left(  {\boldsymbol{\pi}_{0}}%
,{\boldsymbol{\pi}_{1}},{\boldsymbol{\pi}_{2}},\ldots,{\boldsymbol{\pi}_{2n}},{\boldsymbol{\pi}_{2n+1}}\right)  $ of the Markov
process $Q$ is matrix-product, and it is given by
\begin{equation}
{{\boldsymbol{\pi}}_{k+1}} = {{\boldsymbol{\pi}}_{0}}{R_{0}}{R_{1}} \cdots{R_{k}}, \quad 0 \le k \le2n ,
\label{equa-4}%
\end{equation}
where ${{\boldsymbol{\pi}} _{0}}$ is uniquely determined by means of solving the following
system of linear equations
\begin{equation}
{{\boldsymbol{\pi}} _{0}}\left(  {{Q_{0,0}} + {R_{0}}{Q_{1,0}} + {R_{0}}{R_{1}}{Q_{2,0}} +
\cdots+ {R_{0}}{R_{1}} \cdots{R_{2n }}{Q_{2n + 1,0}}} \right)  = 0
\label{equa-1}%
\end{equation}
with the normalized condition
\begin{equation}
{\boldsymbol{\pi} _0} = 1/\left( {e + \sum\limits_{k = 0}^{2n} {{R_0}{R_1} \ldots {R_{k}}e} } \right). \label{equa-2}%
\end{equation}
\end{The}

\textbf{Proof.} From ${\boldsymbol{\pi}  } Q = 0$ and ${\boldsymbol{\pi}  } e
= 1$, we can obtain
\begin{equation}
\left\{
\begin{array}
[c]{l}%
{\boldsymbol{\pi}_{k }}{Q_{k,k+1}} + {\boldsymbol{\pi}_{k+1}}{Q_{k+1,k+1}} = 0,\quad 0 \le k \le 2n,\\
\sum\limits_{k = 0}^{2n + 1} {{\boldsymbol{\pi}_{k}}{Q_{k,0}} = 0,}\\
\label{equa-3} \sum\limits_{k = 0}^{2n + 1} {{\boldsymbol{\pi}_{k}}e = 1.}%
\end{array}
\right.
\end{equation}
Using (\ref{equa-3}), we obtain
\[
{\boldsymbol{\pi}_{1}} = {\boldsymbol{\pi}_{0}}{Q_{0,1}}{\left(  { - {Q_{1,1}^{ - 1}}} \right)  }%
={\boldsymbol{\pi}_{0}}{R_{0}},
\]
\[
{\boldsymbol{\pi}_{2}} = {\boldsymbol{\pi}_{1}}{Q_{1,2}}{\left(  { - {Q_{2,2}^{ - 1}}} \right)  }=
{\boldsymbol{\pi}_{1}}{R_{1}}={\boldsymbol{\pi}_{0}}{R_{0}}{R_{1}},
\]
\[
\vdots
\]
\[
{\boldsymbol{\pi}_{2n + 1}} = {\boldsymbol{\pi}_{2n}}{Q_{2n,2n + 1}}{\left(  { - {Q_{2n + 1,2n + 1}^{ -
1}}} \right)  }={\boldsymbol{\pi}_{2n}}{R_{2n}}={{\boldsymbol{\pi}}_{0}}{R_{0}}{R_{1}} \cdots{R_{2n}}.
\]
Therefore, we obtain
\[
{{\boldsymbol{\pi}}_{k+1}} = {{\boldsymbol{\pi}}_{0}}{R_{0}}{R_{1}} \cdots{R_{k}}, \quad 0 \le k \le2n .
\]
Further, using equations (\ref{equa-4}) and (\ref{equa-3}), we
can obtain the boundary equation (\ref{equa-1}) and the normalized
condition (\ref{equa-2}). This proof is completed. $\square$

Using the stationary probability vector $\boldsymbol{\pi}  $, we can
provide reliability analysis of the PBFT-based blockchain system with repairable voting
nodes as follows:

\textbf{(a) The inherent stationary availability of the PBFT-based blockchain system with repairable voting nodes }is given by
\[{A_2} = 1 - \left( {{\pi _{0,0,n + 1}} + {\pi _{1,0,n + 1}} +  \cdots  + {\pi _{2n,0,n + 1}}} \right) = 1 - \sum\limits_{k = 0}^{2n} {{\pi _{k,0,n + 1}}} .\]

\textbf{(b) The operational stationary availability of the PBFT-based blockchain system with repairable voting nodes }is given by
\[{A_3} = 1 - P_{O} ,\]
where $P_{O}$ represents the stationary probability that the transaction package becomes an orphan block, thus we have
\begin{align*}
P_{O}= \sum_{k=0}^{2 n} \sum_{i=0}^{n+1} \sum_{j=n+1-i} \pi_{k, i, j} .
\end{align*}

Finally, we provide expression for the operational reliability $R_2(t)$ of the PBFT-based blockchain system. To this end, let all the states in the set
\[\left\{ {(0,0,n + 1),(0,1,n), \ldots ,(2n,n+1,0)} \right\}\]
be absorption state ${\nabla _2}$. Then the Markov process $\left\{  {\left(  {N(t),M(t),K(t)} \right)  :t \ge0} \right\}$ operates on a new state space ${\Omega _\nabla } \cup \left\{ {{\nabla _2}} \right\}$, where,
\[
\Omega _\nabla= \mathop    \cup\limits_{k = 0}^{2n + 1} \mathrm{{ Level }}\; k,
\]
for $0 \le k \le 2n+1$,
\begin{align*}
\mathrm{{Level }}\;k =  &  \left\{  {(k,0,0),(k,0,1), \ldots,(k,0,n -
2),(k,0,n - 1),(k,0,n)} \right.  ;\\
&  (k,1,0),(k,1,1), \ldots,(k,1,n - 2),(k,1,n - 1);\\
&  (k,2,0),(k,2,1), \ldots,(k,2,n - 2); \ldots;\\
&  \left.  {(k,n,0)} \right\}  .
\end{align*}

We write the infinitesimal generator of the the Markov process $\left\{ {\left( {N(t),M(t),K(t)} \right):t \ge 0} \right\}$ with the absorption state ${\nabla _2}$ as
\[{Q_{{\nabla _2}}} = \left( {\begin{array}{*{20}{c}}
{{S_\nabla }}&{S_\nabla ^0}\\
0&0
\end{array}} \right).\]
Let $(\phi (0),{\phi _0}(0))$ be the initial probability distribution of the Markov process ${Q_{{\nabla _2}}}$ for ${\phi _0}(0) = 0$ and $\phi (0) = \left( {1,0,0, \ldots ,0} \right)$,
\[{\tau _2} = \inf \left\{ {t \ge 0:M(t)+K(t) = n + 1,\left( {N(0),M(0),K(0)} \right) = \left( {0,0,0} \right)} \right\},\]
and $R_2(t)=P\left\{ {\tau_2  > t}\right\}$, then following theorem provides expression for the operational reliability $R_2(t)$ of the PBFT-based blockchain system.

\begin{The} \label{The-8}
If the initial probability distribution of the Markov process $Q_{\nabla_2}$ is $(\phi (0),{\phi _0}(0))$, then the operational reliability $R_2(t)$ of the PBFT-based blockchain system is given by
\[
R_2(t) = P\left\{ {\tau_2  > t} \right\} = \sum\limits_{k = 0}^{2n+1} {{\phi _k}(t)},
\]
where ${\phi}(t)=({\phi _0}(t),{\phi _1}(t), \ldots, {\phi _{2n+1}}(t))$ satisfy the Chapman-Kolmogorov forward differential equation
\[
\phi '(t) = \phi (t){S_\nabla },
\]
and
\[\phi (t) = \phi (0)\exp \left\{ {{S_\nabla }t} \right\},\]
where $(\phi (0),{\phi _0}(0))=(1,0,0,\ldots ,0)$.
\end{The}

Based on Theorem \ref{The-8}, the average time before the first failure of the PBFT-based blockchain system is
\[{\rm{MTTFF_2 = }}\int_0^\infty  {R_2(t)dt =  - \phi (0)S_\nabla ^{ - 1}e} .\]

\section{Numerical Analysis}\label{sec:numberical analysis}

In this section, we first provide an efficient algorithm for computing the transaction throughput of the PBFT-based blockchain system with repairable voting nodes. Then we use two groups of numerical examples to verify the validity of our theoretical results and to show how some key system parameters influence performance measures of the PBFT-based blockchain system.

\textbf{Group One: The impact of important parameters on transaction throughput $\mathrm{{TH}}$}

Before exploring the impact of some important parameters on the transaction throughput of the PBFT-based blockchain system with repairable voting nodes, it is necessary and useful to provide an approximate algorithm, Algorithm \ref{alg:alg1}, to calculate the transaction throughput, which plays an important role in our numerical examples.
\begin{algorithm*}
  \caption{Approximately computing transaction throughput $\mathrm{{TH}}$ \label{alg:alg1}}
  \KwIn{ The key parameters: $\mu$, $\theta$, $\gamma$, $\beta$, $p$, $\lambda$, $b$, $n$; \\
  \qquad\qquad A controllable accuracy $\varepsilon$  }%
  \KwOut{Transaction throughput $\rm{TH}$ of the PBFT-based blockchain system}
 Determine transition blocks: $\tilde{T}$, $\tilde{S}$ and initial probability vectors: $\tilde{\alpha}$, $\tilde{\omega}$  \;
 Compute the average rates by using ${r_B} = 1/( - \tilde \alpha {\tilde T^{ - 1}}e)$ and ${r_O} = 1/( - \tilde \omega {\tilde S^{ - 1}}e)$ \;
 Based on the obtained rates $r_B$ and $r_O$, determine the transition blocks of the QBD process
\[\left\{B_{1}^{(0)}, A_{2}^{(1)}, A_{0}^{(0)}, A_{2}, A_{1}, A_{0} \right\} ; \] \\
 Use equation (\ref{equa-5}) to compute the rate matrix $R$, stop the iteration if
\[\left\|  R_{\mathbf{n}+1}-R_{\mathbf{n}}\right\|  <\varepsilon ,\]
 and let $R \approx{R_{\mathbf{{n}}}}$ \;
 Solve ${\psi_{0}}$ and ${\psi_{1}}$ by the system equations (\ref{equa-7}) \;
 Compute $\eta _2$ by equation $\eta _2=1-\psi_{0}e$ \;
 Compute the transaction throughput $\rm{TH}$ by equation $\rm{TH}=b\eta _2 {r_B}$ \;
 Output the transaction throughput $\rm{TH}$.
\end{algorithm*}

Next, we use numerical examples to show the impact of some parameters on the transaction throughput of the PBFT-based blockchain system with repairable voting nodes.

\begin{figure}[pbth]
\centering           \includegraphics[width=8cm]{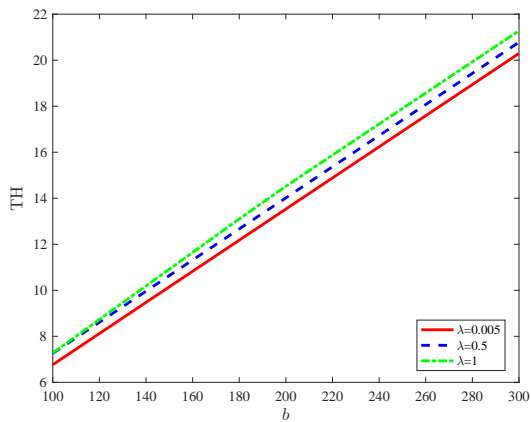}  \caption{$\mathrm{{TH}}$ vs. $b$ and $\lambda$.}%
\label{figure:Fig-13}%
\end{figure}

Firstly, we explore the impact of $\lambda$ and $b$ on $\mathrm{{TH}}$ by means of Algorithm \ref{alg:alg1}. To this end, we take the parameters as follows: $\beta=0.2$, $\gamma=0.5$, $\theta=0.1$, $\mu=0.2$, $n=25$, $p=0.7$, $b\in\left[  {100,300}\right]  $, and $\lambda=0.005,0.1,3$. From Fig. \ref{figure:Fig-13}, we can see that $\mathrm{{TH}}$ increases as $b$ increases, which indicates that the larger the batch size $b$ is, the greater the transaction throughput $\mathrm{{TH}}$ of the PBFT-based blockchain system is. In addition, we can observe that $\mathrm{{TH }}$ increases as $\lambda$ increases. This indicates that as $\lambda$ increases, more and more external transactions arrive at the PBFT-based blockchain system, such that the transaction throughput $\mathrm{{TH }}$ of the PBFT-based blockchain system becomes bigger accordingly. Such numerical results are consistent with our intuitive understanding.

\begin{figure}[pbth]
\centering           \includegraphics[width=8cm]{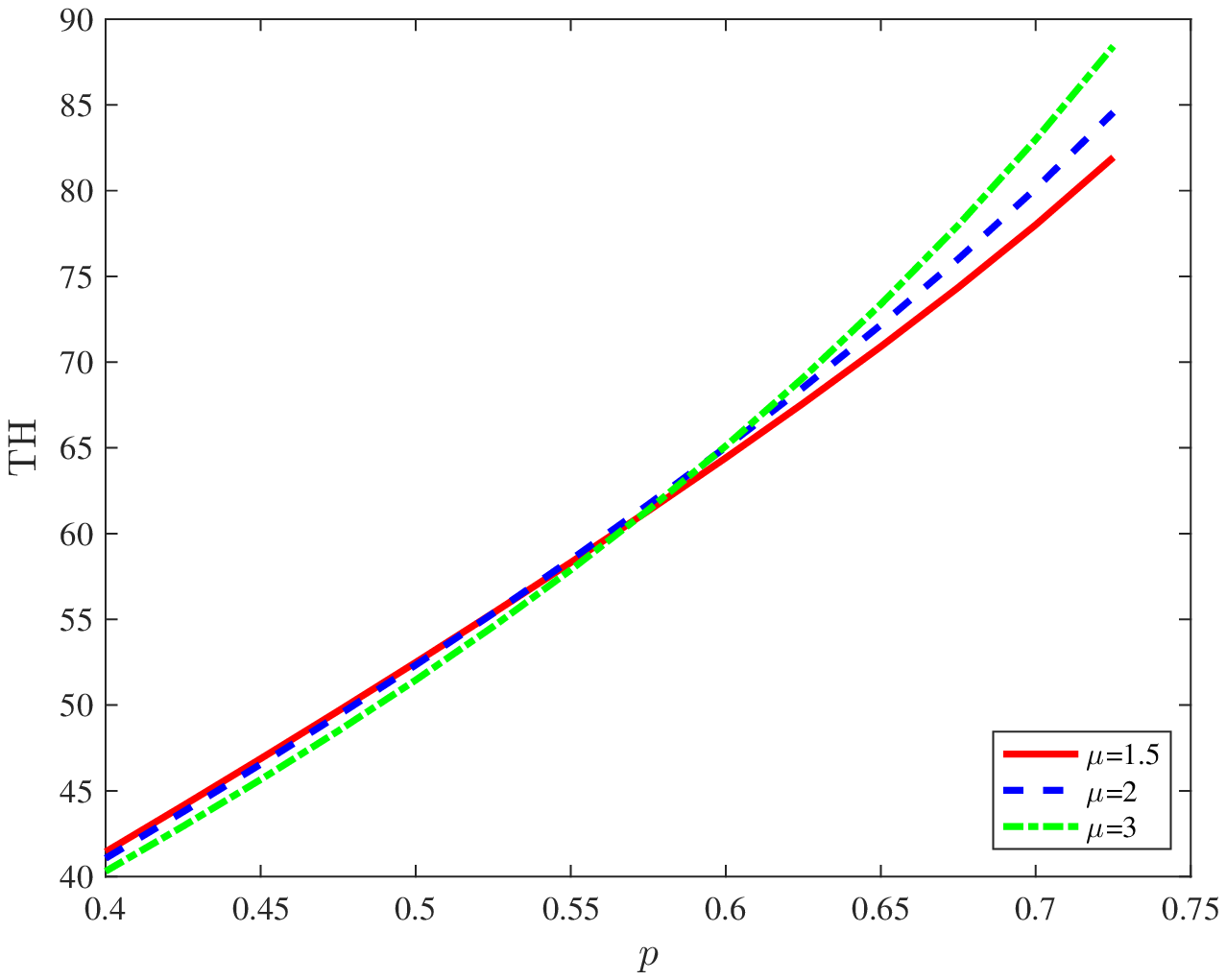}  \caption{$\mathrm{{TH}}$ vs. $p$ and $\mu$.}%
\label{figure:Fig-14}%
\end{figure}

Secondly, we explore the impact of $\mu$ and $p$ on $\mathrm{{TH}}$ by means of Algorithm \ref{alg:alg1}. To this end, we take the parameters as follows: $\beta=3$, $\gamma=5$, $\lambda=2$, $\theta=2$, $n=25$, $b=100$, $p\in\left[  {0.4,0.725}\right]  $, and $\mu=1.5,2,3$. From Fig. \ref{figure:Fig-14}, we can see that $\mathrm{{TH}}$ increases as $p$ increases, which shows that the higher probability of each voting node approving a transaction package, the greater the transaction throughput of the PBFT-based blockchain system is. In addition, we can see that there exists a ${p_0}$, and when $p<{p_0}$, the $\mathrm{{TH}}$ decreases as $\mu$ increases; when $p>{p_0}$, the $\mathrm{{TH}}$ increases as $\mu$ increases. This result indicates that as $\mu$ increases, more and more repaired nodes enter the network, which increases the probability that the PBFT-based blockchain system finds the judgment condition of the voting process. However, the smaller $p$ can increase the chance of determining a transaction package as an orphan block and then decrease the transaction throughput; While the larger $p$ can increase the chance of determining a transaction package as a block, thereby increasing the transaction throughput. Such numerical results are consistent with our intuitive understanding.
\begin{figure}[tbph]
\centering           \includegraphics[width=8cm]{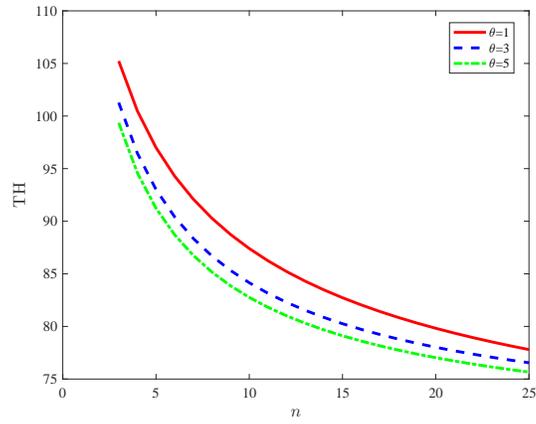}  \caption{$\mathrm{{TH}}$ vs. $n$ and $\theta$.}%
\label{figure:Fig-15}%
\end{figure}

Finally, we explore the impact of $\theta$ and $n$ on $\mathrm{{TH}}$ by means of Algorithm \ref{alg:alg1}. To this end, we take the parameters as follows: $\beta=3$, $\gamma=5$, $\lambda=2$, $\mu=2$, $p=0.68$, $b=100$, $n \in\left[  {3,25} \right]  $, and $\theta= 1,3,5$. From Fig. \ref{figure:Fig-15}, we observe that $\mathrm{{TH}}$ decreases as $n$ increases, which indicates that the increase of $n$ decreases the probability that the PBFT-based blockchain system arrives at the judgment condition, such that the transaction throughput $\mathrm{{TH}}$ decreases. That is, if we aim to pursue the high transaction throughput of a PBFT-based blockchain system, we need a smaller number of total voting nodes. However, if the majority of these nodes are Byzantine, the security of the PBFT-based blockchain system may be compromised. This fact means that we sometimes have to sacrifice the throughput to secure the PBFT-based blockchain system. In addition, we can see that the transaction throughput $\mathrm{{TH}}$ decreases as $\theta$ increases. This result indicates that as $\theta$ increases, more and more failed nodes leave the PBFT-based blockchain system, which can decrease the probability that the PBFT-based blockchain system arrives at the judgment condition, such that the transaction throughput $\mathrm{{TH}}$ decreases. Such numerical results are also consistent with our intuitive understanding.

In what follows, Table \ref{tab:addlabel} shows the orders of some important sub-matrices with different $b$ and $n$. It can be seen that the orders of sub-matrices expand rapidly under the influence of the Kronecker operators and $n$. Such sub-matrices with higher order limit the ability to calculate the transaction throughput $\mathrm{{TH}}$. Therefore, our approximate algorithm provides an effective and convenient method to calculate transaction throughput, so this approximate algorithm has more general applicability and universality to the study of PBFT-based blockchain systems.
\begin{table}[htbp]
  \centering
  \caption{The orders of important sub-matrices with different $b$ and $n$}
 \begin{tabular}{|c|c|c|c|}
\toprule values of $n$ & $\tilde{S}$ or $\tilde{T}$ & ${\tilde{S}\oplus\tilde{T}-\lambda I}$  & $A_1$\\ \hline
    2     & $31\times31$    & $961\times961$ & $961b \times 961b$ \\
    \hline
    3     & $71\times71$    & $5041\times5041$ & $5041b \times 5041b$ \\
    \hline
    4     & $136\times136$   & $18496\times18496$ & $18496b \times 18496b$\\
    \hline
    5     & $231\times231$   & $53361\times53361$ & $53361b \times 53361b$ \\
    \bottomrule
    \end{tabular}%
  \label{tab:addlabel}%
\end{table}

\textbf{Group Two: The impact of important parameters on the availabilities }

\begin{figure}[ptbh]
\centering     \subfigure[${A_{1}}$ vs. $n$ and $\mu$]  {\includegraphics[width=7cm]{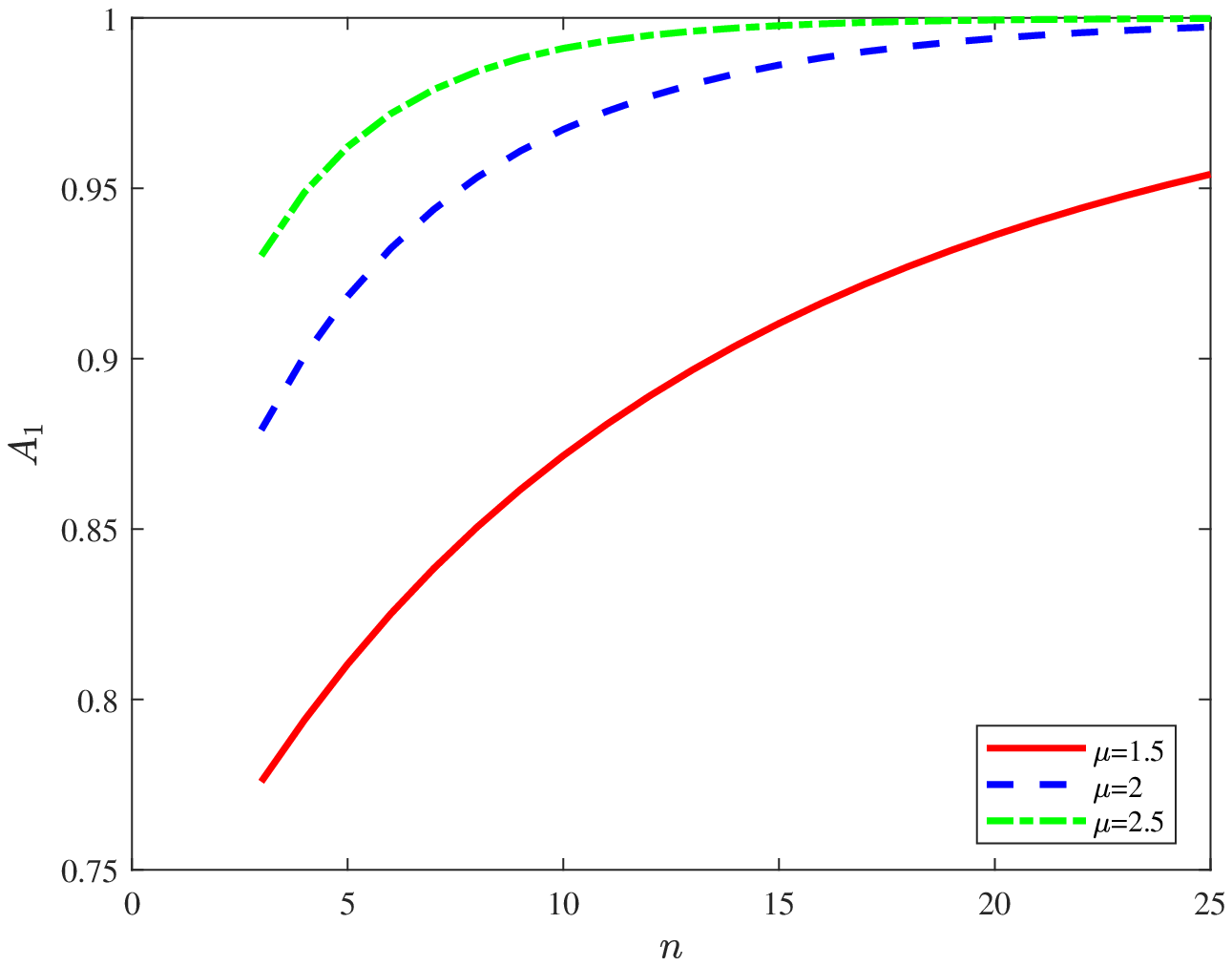} \label{figure16a}
}
\subfigure[${A_{1}}$ vs. $n$ and $\theta$]  { \includegraphics[width=7cm]{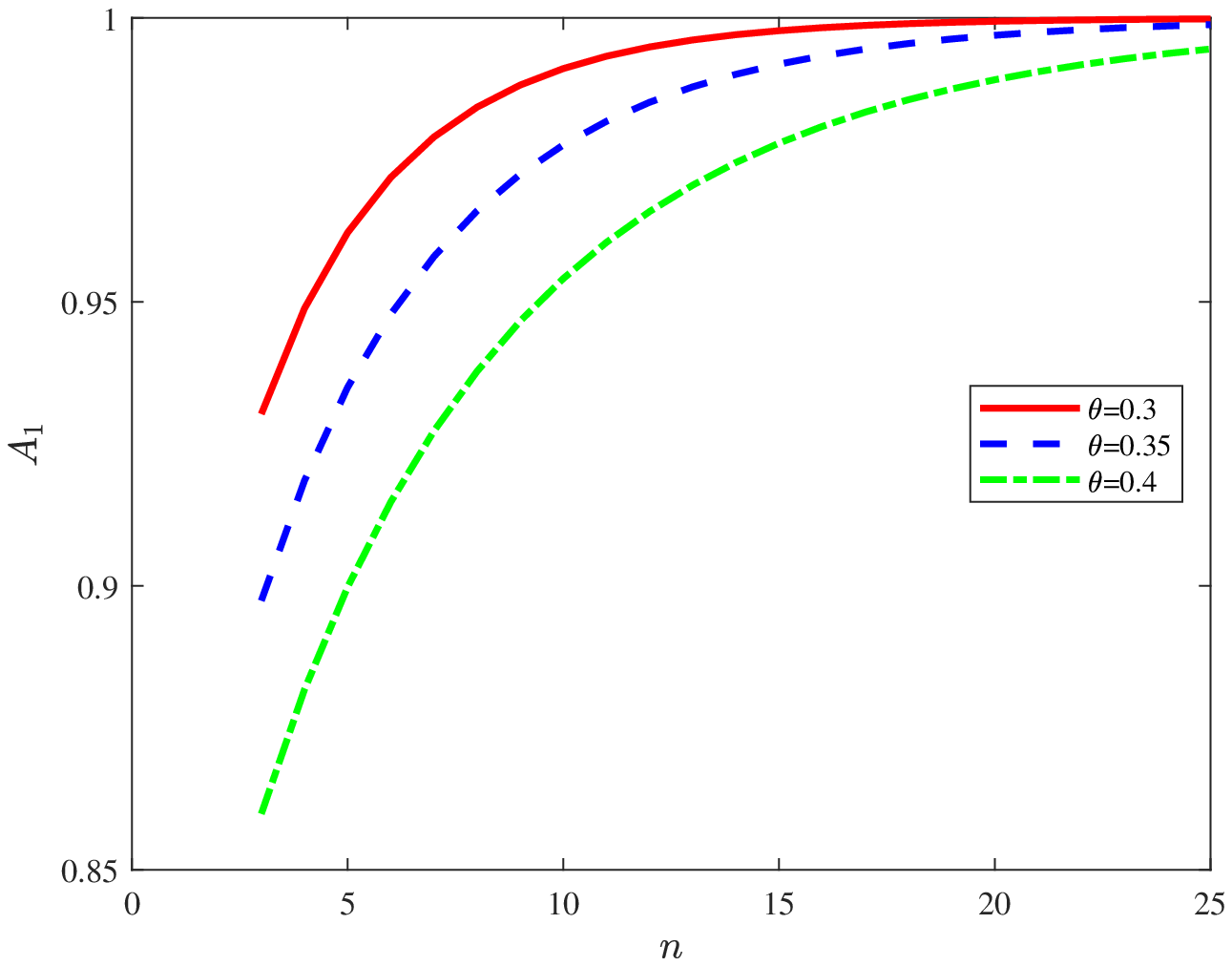} \label{figure16b} }  \caption{$A_{1}$ vs. three parameters $n$, $\mu$ and $\theta$.}%
\label{figure:Fig-16}%
\end{figure}

\begin{figure}[ptbh]
\centering     \subfigure[${A_{2}}$ vs. $n$ and $\mu$]  {\includegraphics[width=7cm]{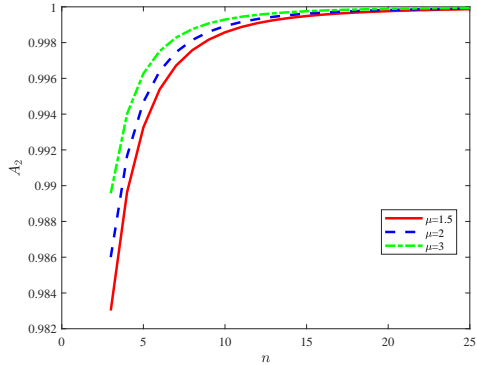} \label{figure17a}
}
\subfigure[${A_{2}}$ vs. $n$ and $\theta$]  { \includegraphics[width=7cm]{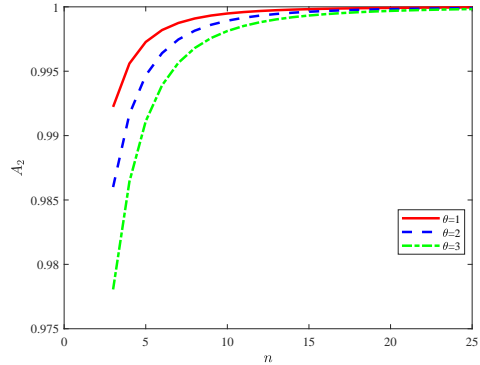} \label{figure17b} }  \caption{$A_{2}$ vs. three parameters $n$, $\mu$ and $\theta$.}%
\label{figure:Fig-17}%
\end{figure}

In this part, we explore the impact of the parameters $n$, $\mu$ and $\theta$ on the availabilities ${A_{1}}$ and ${A_{2}}$. For Fig. \ref{figure16a}, we take the parameters as follows:
$\theta=0.5$, $n\in\left[  {3,25}\right]  $ and $\mu=1.5,2,2.5$; For Fig. \ref{figure16b}, we take the parameters as follows: $\mu=1.5$, $n \in\left[  {3,25} \right]  $, and $\theta= 0.3,0.35,0.4$; For Fig. \ref{figure17a}, we take the parameters as follows:
$\beta=3$, $\theta=2$, $\gamma=10$, $p=0.7$, $n\in\left[  {3,25}\right]  $ and $\mu=1.5,2,3$;
For Fig. \ref{figure17b}, we take the parameters as follows: $\beta=3$, $\mu=2$, $\gamma= 10$, $p=0.7$, $n \in\left[  {3,25} \right]  $, and $\theta= 1,2,3$.

From Fig. \ref{figure:Fig-16} and Fig. \ref{figure:Fig-17}, we can see that the stationary availabilities $A_1$ and $A_2$ increase as $n$ increases, which indicates that the total voting nodes $N=3n+1$ of the PBFT-based blockchain system with repairable voting nodes can increase the stationary availabilities $A_1$ and $A_2$. In other words, the more voting nodes, the higher the stationary availability of the PBFT-based blockchain system is. Also, it can be seen from Fig. \ref{figure:Fig-16} and Fig. \ref{figure:Fig-17} that when $n$ increases to a certain value, the increase of the stationary availability $A_1$ or $A_2$ is no longer obvious. On the other hand, Fig. \ref{figure16a} and Fig. \ref{figure17a} show that the stationary availability $A_1$ (or $A_2$) increases as $\mu$ increases and indicate that the stationary availability $A_1$ (or $A_2$) decreases as $\theta$ increases. These numerical results indicate that when the total number of nodes $N=3n+1$ is constant, more and more repaired nodes enter the PBFT-based blockchain system as $\mu$ increases, which can increase the stationary availability $A_1$ (or $A_2$). Instead, more and more failed nodes leave the PBFT-based blockchain system as $\theta$ increases, which can decrease the stationary availability $A_1$ (or $A_2$). Such numerical results are consistent with our intuitive understanding.

Finally, from Groups One and Two of the numerical examples, we can observe that when the total number of voting nodes is constant, the number of voting nodes in the PBFT-based blockchain system affects the throughput, security, and reliability of the PBFT-based blockchain system. That is, the greater the number of voting nodes is, the higher the availability and security of the PBFT-based blockchain system are, but the lower the throughput is. This result shows again that high throughput, high security, and high availability do not coexist. In this case, we sometimes have to sacrifice the throughput to ensure the security and availability of the PBFT-based blockchain system.

\section{Concluding Remarks}\label{sec:concluding}

In this paper, we propose a new PBFT to generalize the ordinary PBFT by introducing a process in which the voting nodes in the PBFT-based blockchain system can fail and be repaired. Specifically, the voting nodes in the new PBFT-based blockchain system may leave the PBFT network while some repaired nodes can re-enter the PBFT network again. In this case, the number of working voting nodes in the new PBFT-based blockchain system is variable, but the total number of voting nodes must be constant. For this PBFT-based blockchain system with repairable voting nodes, we provide a large-scale Markov modeling method to analyze the PBFT-based blockchain system. Also, we provide performance analysis for the PBFT-based blockchain system with repairable voting nodes. At the same time, we perform a reliability analysis for the PBFT-based blockchain system with repairable voting nodes and propose some reliability measures, including availability, reliability, and the average time before the first failure. Finally, we use numerical examples to verify the validity of our theoretical results and indicate how some key system parameters influence the performance measures of the PBFT-based blockchain system with repairable voting nodes.

We further develop the research lines that apply the Markov process and queueing theory to provide the performance analysis of blockchain systems under different consensus \cite{Li:2018, Li:2019, Song:2022, Chang:2022}, where the computation is supported by the RG-factorization technique \cite{Li:2010}. Based on this, we are optimistic to believe that the methodology and results given in this paper can be applied to dealing with more general PBFT-based blockchain systems in practice, and develop some effective algorithms to enhance the throughput, security, and reliability of the PBFT-based blockchain systems from the purpose of many actual uses.

Along these research lines, we will continue our future research in the following directions:

---When the lifetime or/and repair time of a voting node is of phase type, an interesting future research is to focus on finding efficient algorithms to deal with the multi-dimensional Markov processes with a block structure corresponding to the PBFT-based blockchain systems.

---Setting up reward functions with respect to cost structure, transaction fee, block reward, blockchain security, and so forth. It is very interesting in our future study to develop stochastic optimization, Markov decision processes, and stochastic game models in the study of PBFT-based blockchain systems.

---In the direction of stochastic optimization, dynamic control, and Markov decision processes for the PBFT-based blockchain systems, develop efficient algorithms for dealing with the optimal policy from the throughput, security, and reliability.
\section*{Acknowledgment}

This work was supported by the National Natural Science Foundation of China
under grant No. 71932002.

\section*{Two Appendixes}
We provide two appendixes for the two infinitesimal generators $\Theta$ given in Subsection \ref{subsec:orphan-generated time} and $Q$ given in Subsection \ref{subsec:complex reliability}. Our purpose is to increase the readability of the main paper.

\textbf{Appendix A: The infinitesimal generator $\Theta$}
\[
\Theta=\left(
\begin{array}
[c]{cc}%
0 & 0\\
{{S^{0}}} & S
\end{array}
\right)  ,
{S^{0}}+S\mathbf{{e}}=0,
\]
where,
\[
{S^{0}}=\left(  {%
\begin{array}
[c]{c}%
{S_{0}^{0}}\\
{S_{1}^{0}}\\
\vdots\\
{S_{2n-1}^{0}}\\
{S_{2n}^{0}}%
\end{array}
}\right)  ,S_{k}^{0}=\left(  {%
\begin{array}
[c]{c}%
{S_{k,0}^{0}}\\
{S_{k,1}^{0}}\\
\vdots\\
{S_{k,n-1}^{0}}\\
{S_{k,n}^{0}}%
\end{array}
}\right)  ,0\leq k\leq2n;
\]%
\[
S_{k,i}^{0}={\left(  {%
\begin{array}
[c]{c}%
0\\
\vdots\\
0\\
{(N-k-n)(\theta+\gamma q)}%
\end{array}
}\right)  _{(n+1-i)\times1}},\quad 0\leq i\leq n-1,
\]%
\[
S_{k,n}^{0}=(N-k-n)(\theta+\gamma q);
\]%
\[
S=\left(  {%
\begin{array}
[c]{cccccc}%
{{S_{0,0}^{(k)}}} & {{S_{0,1}^{(k)}}} &  &  &  & \\
& {{S_{1,1}^{(k)}}} & {{S_{1,2}^{(k)}}} &  &  & \\
&  & {{S_{2,2}^{(k)}}} & {{S_{2,3}^{(k)}}} &  & \\
&  &  & \ddots & \ddots & \\
&  &  &  & {{S_{2n-1,2n-1}^{(k)}}} & {{S_{2n-1,2n}^{(k)}}}\\
&  &  &  &  & {{S_{2n,2n}^{(k)}}}%
\end{array}
}\right)  ,
\]
for $0\leq k\leq2n-1$,
\[
{S_{k,k+1}}=\left(  {%
\begin{array}
[c]{cccc}%
{{F_{0,0}^{(k)}}} &  &  & \\
& {{F_{1,1}^{(k)}}} &  & \\
&  & \ddots & \\
&  &  & {{F_{n,n}^{(k)}}}%
\end{array}
}\right)  ,
\]
for $0\leq i\leq n$,
\[
{F_{i,i}^{(k)}}={\left(  {%
\begin{array}
[c]{cccc}%
{(N-k-i)\gamma p} &  &  & \\
& {(N-k-i-1)\gamma p} &  & \\
&  & \ddots & \\
&  &  & {(N-k-n)\gamma p}%
\end{array}
}\right)  _{(n+1-i)\times(n+1-i)}};
\]
for $0\leq k\leq2n-1$,
\[
{S_{k.k}}=\left(  {%
\begin{array}
[c]{ccccc}%
{{G_{0,0}^{(k)}}} & {{G_{0,1}^{(k)}}} &  &  & \\
& {{G_{1,1}^{(k)}}} & {{G_{1,2}^{(k)}}} &  & \\
&  & \ddots & \ddots & \\
&  &  & {{G_{n-1,n-1}^{(k)}}} & {{G_{n-1,n}^{(k)}}}\\
&  &  &  & {{G_{n,n}^{(k)}}}%
\end{array}
}\right),
\]
for $0\leq i\leq n-1$,
\[
{G_{i,i+1}^{(k)}}={\left(  {%
\begin{array}
[c]{cccc}%
{(N-k-i)\gamma q} &  &  & \\
& {(N-k-i-1)\gamma q} &  & \\
&  & \ddots & \\
&  &  & {(N-k-n+1)\gamma q}\\
&  &  & 0
\end{array}
}\right)  _{(n+1-i)\times(n-i)}},
\]%
\[
{G_{i,i}^{(k)}}={\left(  {%
\begin{array}
[c]{ccccc}%
{{c_{k,i,0}}} & {(N-k-i)\theta} &  &  & \\
\mu & {{c_{k,i,1}}} & {(N-k-i-1)\theta} &  & \\
& \ddots & \ddots & \ddots & \\
&  & {(n-i-1)\mu} & {{c_{k,i,n-i-1}}} & {(N-k-n+1)\theta}\\
&  &  & {(n-i)\mu} & {{c_{k,i,n-i}}}%
\end{array}
}\right)  _{(n+1-i)\times(n+1-i)}},
\]%
\[
{c_{k,i,m}}=-\left[  {(N-k-i-m)(\theta+\gamma)+m\mu}\right]  ,{0}\leq m\leq n-i;
\]%
\[
{G_{n,n}^{(k)}}=-(N-k-n)(\theta+\gamma);
\]%
\[
{S_{2n,2n}} = \left(  {%
\begin{array}
[c]{ccccc}%
{{H_{0,0}}} & {{H_{0,1}}} & {} & {} & {}\\
{} & {{H_{1,1}}} & {{H_{1,2}}} & {} & {}\\
{} & {} & \ddots & \ddots & {}\\
{} & {} & {} & {{H_{n - 1,n - 1}}} & {{H_{n - 1,n}}}\\
{} & {} & {} & {} & {{H_{n,n}}}%
\end{array}
} \right)  ,
\]
for $0 \le i \le n - 1$,
\[
{H_{i,i + 1}} = {\left(  {%
\begin{array}
[c]{cccc}%
{(n + 1 - i)\gamma q} & {} & {} & {}\\
{} & {(n - i)\gamma q} & {} & {}\\
{} & {} & \ddots & {}\\
{} & {} & {} & {2\gamma q}\\
{} & {} & {} & 0
\end{array}
} \right)  _{(n + 1 - i) \times(n - i)}},
\]%
\[
{H_{i,i}} = {\left(  {%
\begin{array}
[c]{ccccc}%
{{f_{i,0}}} & {(n + 1 - i)\theta} & {} & {} & {}\\
\mu & {{f_{i,1}}} & {(n - i)\theta} & {} & {}\\
{} & \ddots & \ddots & \ddots & {}\\
{} & {} & {(n - i - 1)\mu} & {{f_{i,n - i - 1}}} & 2\theta\\
{} & {} & {} & {(n - i)\mu} & {{f_{i,n - i}}}%
\end{array}
} \right)  _{(n + 1 - i) \times(n + 1 - i)}},
\]
\[{f_{i,m}} = - \left[  {(n+1 - i - m)(\theta+ \gamma q) + m\mu} \right]
,\mathrm{{ 0}} \le m \le n - i-1,\]
\[
f_{i,n - i}=-\left[  {(N-k-n)\gamma p+(n-i)\mu}\right];
\]
\[
{H_{n,n}} = - (\theta+ \gamma q).
\]
\textbf{Appendix B: The infinitesimal generator $Q$}
\[
Q = \left(  {%
\begin{array}
[c]{cccccc}%
{{Q_{0,0}}} & {{Q_{0,1}}} & {} & {} & {} & {}\\
{{Q_{1,0}}} & {{Q_{1,1}}} & {{Q_{1,2}}} & {} & {} & {}\\
{{Q_{2,0}}} & {} & {{Q_{2,2}}} & {{Q_{2,3}}} & {} & {}\\
\vdots & {} & {} & \ddots & \ddots & {}\\
{{Q_{2n,0}}} & {} & {} & {} & {{Q_{2n,2n}}} & {{Q_{2n,2n + 1}}}\\
{{Q_{2n + 1,0}}} & {} & {} & {} & {} & {{Q_{2n + 1,2n + 1}}}%
\end{array}
} \right)  ,
\]
where,
\[
{Q_{2n + 1,0}} = {\left(  {%
\begin{array}
[c]{cc}%
{{J_{0,0}}} & {}\\
{{J_{1,0}}} & {}\\
\vdots & {}\\
{{J_{n,0}}} & {}%
\end{array}
} \right)  _{\left(  {\sum\limits_{i = 1}^{n + 1} i } \right)  \times\left(
{\sum\limits_{i = 1}^{n + 2} i } \right)  }},
\]
\[
{J_{i,0}} = {\left(  {%
\begin{array}
[c]{cc}%
\beta & {}\\
\beta & {}\\
\vdots & {}\\
\beta & {}%
\end{array}
} \right)  _{\left(  {n + 1 - i} \right)  \times(n + 2)}},\quad0 \le i \le n;
\]
\[
{Q_{2n + 1,2n + 1}} = \left(  {%
\begin{array}
[c]{ccccc}%
{{K_{0,0}}} & {{K_{0,1}}} & {} & {} & {}\\
{} & {{K_{1,1}}} & {{K_{1,2}}} & {} & {}\\
{} & {} & \ddots & \ddots & {}\\
{} & {} & {} & {{K_{n - 1,n - 1}}} & {{K_{n - 1,n}}}\\
{} & {} & {} & {} & {{K_{n,n}}}%
\end{array}
} \right)  ,
\]
for $0 \le i \le n-1$,
\[
{K_{i,i + 1}} = {\left(  {%
\begin{array}
[c]{cccc}%
{(n - i)\gamma q} & {} & {} & {}\\
{} & {(n - i - 1)\gamma q} & {} & {}\\
{} & {} & \ddots & {}\\
{} & {} & {} & {\gamma q}\\
{} & {} & {} & 0
\end{array}
} \right)  _{(n + 1 - i) \times(n - i)}},
\]%
\[
{K_{i,i}}={\left(  {%
\begin{array}
[c]{ccccc}%
{{a_{i,0}}} & {(n-i)\theta} &  &  & \\
\mu & {{a_{i,1}}} & {(n-i-1)\theta} &  & \\
& \ddots & \ddots & \ddots & \\
&  & {(n-i-1)\mu} & {{a_{i,n-i-1}}} & \theta\\
&  &  & {(n-i)\mu} & {{a_{i,n-i}}}%
\end{array}
}\right)  _{(n+1-i)\times(n+1-i)}};
\]
\[
{a_{i,m}} = - \left[  {(n - i - m)(\theta+ \gamma q) + m\mu+ \beta} \right]
,\quad\mathrm{{ 0}} \le m \le n - i,
\]
\[{K_{n,n}} = - \beta;\]
for $1\leq k\leq2n$,
\[
{Q_{k,0}}={\left(  {%
\begin{array}
[c]{cc}%
{{A_{0,0}}} & \\
{{A_{1,0}}} & \\
\vdots & \\
{{A_{n+1,0}}} &
\end{array}
}\right)  _{\left(  {\sum\limits_{i=1}^{n+2}i}\right)  \times\left(
{\sum\limits_{i=1}^{n+2}i}\right)  }},
\]
\[
{A_{i,0}}={\left(  {%
\begin{array}
[c]{cc}%
0 & \\
\vdots & \\
0 & \\
\beta &
\end{array}
}\right)  _{(n+2-i)\times(n+2)}},\quad0\leq i\leq n,
\]%
\[
{A_{n+1,0}}={\left(  {\beta,0,0,\ldots,0}\right)  _{1\times(n+2)}};
\]
for $0\leq k\leq2n-1$,
\[
{Q_{k,k+1}}=\left(  {%
\begin{array}
[c]{ccccc}%
{{B_{0,0}^{(k)}}} &  &  &  & \\
& {{B_{1,1}^{(k)}}} &  &  & \\
&  & \ddots &  & \\
&  &  & {{B_{n,n}^{(k)}}} & \\
&  &  &  & 0
\end{array}
}\right)  ,
\]
for $0\leq i\leq n$,
\[
{B_{i,i}^{(k)}}={\left(  {%
\begin{array}
[c]{ccccc}%
{(N-k-i)\gamma p} &  &  &  & \\
& {(N-k-i-1)\gamma p} &  &  & \\
&  & \ddots &  & \\
&  &  & {(N-k-n)\gamma p} & \\
&  &  &  & 0
\end{array}
}\right)  _{(n+2-i)\times(n+2-i)}};
\]%
\[
{Q_{2n,2n + 1}} = \left(  {%
\begin{array}
[c]{cccc}%
{{C_{0,0}}} & {} & {} & {}\\
{} & {{C_{1,1}}} & {} & {}\\
{} & {} & \ddots & {}\\
{} & {} & {} & {{C_{n,n}}}\\
{} & {} & {} & 0
\end{array}
} \right)  ,
\]
where,
\[
{C_{i,i}} = {\left(  {%
\begin{array}
[c]{cccc}%
{(n + 1 - i)\gamma p} & {} & {} & {}\\
{} & {(n - i)\gamma p} & {} & {}\\
{} & {} & \ddots & {}\\
{} & {} & {} & {\gamma p}\\
{} & {} & {} & 0
\end{array}
} \right)  _{(n + 2 - i) \times(n + 1 - i)}},\quad0 \le i \le n;
\]%
\[
{Q_{0,0}}=\left(  {%
\begin{array}
[c]{cccccc}%
{{D_{0,0}}} & {{D_{0,1}}} &  &  &  & \\
{{D_{1,0}}} & {{D_{1,1}}} & {{D_{1,2}}} &  &  & \\
{{D_{2,0}}} &  & {{D_{2,2}}} & {{D_{2,3}}} &  & \\
\vdots &  &  & \ddots & \ddots & \\
{{D_{n,0}}} &  &  &  & {{D_{n,n}}} & {{D_{n,n+1}}}\\
{{D_{n+1,0}}} &  &  &  &  & {{D_{n+1,n+1}}}%
\end{array}
}\right)  ,
\]
\[
{D_{n+1,0}}={\left(  {\beta,0,\ldots,0}\right)  _{1\times(n+2)}},{D_{n+1,n+1}%
}=-\beta,
\]%
\[
{D_{i,0}}={\left(  {%
\begin{array}
[c]{cc}%
0 & \\
\vdots & \\
0 & \\
\beta &
\end{array}
}\right)  _{(n+2-i)\times(n+2)}},1\leq i\leq n,
\]
for $0\leq i\leq n$,
\[
{D_{i,i+1}}={\left(  {%
\begin{array}
[c]{cccc}%
{(N-i)\gamma q} &  &  & \\
& {(N-i-1)\gamma q} &  & \\
&  & \ddots & \\
&  &  & {(N-n)\gamma q}\\
&  &  & 0
\end{array}
}\right)  _{(n+2-i)\times(n+1-i)}};
\]
\[
{b_{i,m}}=-\left[  {(N-i-m)(\theta+\gamma)+m\mu}\right]  ,\mathrm{{0}}\leq
m\leq n-i,
\]%
\[
{d_{i}}={(n+1-i)\mu},\quad 0\leq i\leq n,
\]%
\[
{D_{0,0}}={\left(  {%
\begin{array}
[c]{ccccc}%
{-{b_{0,0}}} & {N\theta} &  &  & \\
\mu & {-{b_{0,1}}} & {(N-1)\theta} &  & \\
& \ddots & \ddots & \ddots & \\
&  & {n\mu} & {-{b_{0,n}}} & {(N-n)\theta}\\
\beta &  &  & {(n+1)\mu} & {-d_{0}-\beta}%
\end{array}
}\right)  _{(n+2)\times(n+2)}},
\]
for $1\leq i\leq n$,
\[
{D_{i,i}}={\left(  {%
\begin{array}
[c]{ccccc}%
{-{b_{i,0}}} & {(N-i)\theta} &  &  & \\
\mu & {-{b_{i,1}}} & {(N-i-1)\theta} &  & \\
& \ddots & \ddots & \ddots & \\
&  & {(n-i)\mu} & {-{b_{i,n-i}}} & {(N-n)\theta}\\
&  &  & {d_{i}} & {-d_{i}-\beta}%
\end{array}
}\right)  _{(n+2-i)\times(n+2-i)}};
\]
for $1\leq k\leq2n$,
\[
{Q_{k,k}}=\left(  {%
\begin{array}
[c]{ccccccc}%
{{E_{0,0}^{(k)}}} & {{E_{0,1}^{(k)}}} &  &  &  &  & \\
& {{E_{1,1}^{(k)}}} & {{E_{1,2}^{(k)}}} &  &  &  & \\
&  & {{E_{2,2}^{(k)}}} & {{E_{2,3}^{(k)}}} &  &  & \\
&  &  & \ddots & \ddots &  & \\
&  &  &  & {{E_{n,n}^{(k)}}} & {{E_{n,n+1}^{(k)}}} & \\
&  &  &  &  & {{E_{n+1,n+1}^{(k)}}} &
\end{array}
}\right)  ,
\]
\[
{E_{n+1,n+1}^{(k)}}=-\beta,
\]
for $0\leq i\leq n$,
\[
{E_{i,i+1}^{(k)}}={\left(  {%
\begin{array}
[c]{cccc}%
{(N-k-i)\gamma q} &  &  & \\
& {(N-k-i-1)\gamma q} &  & \\
&  & \ddots & \\
&  &  & {(N-k-n)\gamma q}\\
&  &  & 0
\end{array}
}\right)  _{(n+2-i)\times(n+1-i)}},
\]%
\[
{E_{i,i}^{(k)}} = {\left(  {%
\begin{array}
[c]{ccccc}%
{ - {c_{k,i,0}}} & {(N - k - i)\theta} & {} & {} & {}\\
\mu & { - {c_{k,i,1}}} & {(N - k - i - 1)\theta} & {} & {}\\
{} & \ddots & \ddots & \ddots & {}\\
{} & {} & {(n - i)\mu} & { - {c_{k,i,n - i}}} & {(N - k - n)\theta}\\
{} & {} & {} & {d_{i} } & { -d_{i}-\beta}%
\end{array}
} \right)  _{(n + 2 - i) \times(n + 2 - i)}},
\]
\[
{c_{k,i,m}}=-\left[  {(N-k-i-m)(\theta+\gamma)+m\mu}\right]  ,\mathrm{{0}}\leq
m\leq n-i.
\]%

\end{document}